\documentclass[11pt]{article}

\usepackage[a4paper, margin=1in]{geometry}
\usepackage{graphicx}
\usepackage{natbib}
\usepackage{setspace}

\usepackage{appendix}

\usepackage{amssymb}
\usepackage{amsmath,amsthm}

\usepackage[colorlinks=True,citecolor=blue]{hyperref}
\usepackage{url}
\usepackage{cleveref}
\usepackage{titling}

\newtheorem{theorem}{Theorem}
\newtheorem{lemma}[theorem]{Lemma}
\newtheorem{definition}{Definition}

\newcommand{\limd}[1]{\underset{#1}{\lim }\,}
\renewcommand{\Pr}[2][{}]{\mathbb{P}_{#1}\left(#2\right)}

\title{The marginal majority effect: when social influence produces lock-in}
\begin{document}
\onehalfspacing
\date{}

\author{Alexandros Gelastopoulos, Pantelis P. Analytis, Ga\"el Le Mens,\\ and Arnout van de Rijt}

\maketitle

\begin{abstract}
    People are influenced by the choices of others, a phenomenon observed across contexts in the social and behavioral sciences. Social influence can lock in an initial popularity advantage of an option over a higher quality alternative. Yet several experiments designed to enable social influence have found that social systems self-correct rather than lock-in. Here we identify a behavioral phenomenon that makes inferior lock-in possible, which we call the `marginal majority effect': A discontinuous increase in the choice probability of an option as its popularity exceeds that of a competing option. We demonstrate the existence of marginal majority effects in several recent experiments and show that lock-in always occurs when the effect is large enough to offset the quality effect on choice, but rarely otherwise. Our results reconcile conflicting past empirical evidence and connect a behavioral phenomenon to the possibility of social lock-in.
\end{abstract}


Whether it is about choosing a party to vote for \citep{simon1954bandwagon}, deciding whether to adopt an innovation \citep{bass1969new}, selecting music to listen to \citep{salganik2006experimental} or whether to get vaccinated during a pandemic \citep{rabb2022influence}, people tend to select options that many other people have selected before them \citep{asch1955opinions,arthur1989competing,brock2001discrete,kohler1997learning}. Work on network externalities, information cascades, and herding in economics \citep{david1985clio,arthur1989competing,banerjee1992simple,bikhchandani1992theory}, on normative conformism in anthropology and biology \citep{boyd1982cultural,boyd1988culture}, and on status-quality dissociation in sociology \citep{lynn2009sociological} has pointed out that when social influence is strong enough, it can trigger a reinforcing feedback loop that drives popular options to become even more popular, leading to path dependence and potentially lock-in on inferior outcomes. This hypothesis has found support in historical empirical evidence \citep{david1985clio,cusumano1992strategic,akerlof2018persistence} and more recently in large scale experimental studies \citep{centola2018experimental,macy2019opinion}.

Yet the same hypothesis has also been refuted in experiments across the social and behavioral sciences that were designed to enable social influence. In the classic social conformity paradigm in psychology, social influence is too weak to sustain a majority of responses for the wrong answer \citep{asch1955opinions, efferson2008conformists,van2019self}. In experiments where the information cascade model predicts that lock-in should occur, majorities favoring the incorrect choice tend to self-correct \citep{goeree2007self}. Similarly, experiments on platform choice in economics show that people readily switch to a superior platform when it becomes available, overcoming network effects \citep{hossain2009quest}. In sociological and political science experiments, social influence is not sufficiently strong for popularizing mediocre songs, less fit presidential candidates, or bad art \citep{salganik2006experimental, salganik2008leading, van2019self}. And in an experiment on crowd wisdom in the answering of knowledge questions, some questions frequently produced lock-in on the incorrect answer but other questions never did \citep{frey2021social}. This body of evidence suggests that in certain settings, even under conditions of substantial social influence, lock-in might still be impossible; even if history was re-run a large number of times, inferior lock-in would never take place. 

The question of when inferior lock-in is likely to occur has received little attention so far. Although it is conceivable that the difference in quality or inherent appeal of the competing options is a strong determinant of a possible lock-in, it is not clear what aspects of the way people respond to social influence are also relevant. The present paper extends a theoretical framework for binary choice under social influence \citep{simon1954bandwagon,schelling1978micromotives,granovetter1978threshold,boyd1988culture,arthur1989competing,lopez2008social} and identifies a condition that is sufficient for inferior lock-in to be possible. This condition relates to the tendency of people to follow the option a majority of others have chosen, regardless of the size of this majority. More precisely, we define the \textit{marginal majority effect} as the increase in the probability that an individual chooses a certain option when this option becomes marginally more popular than its competition. We show theoretically that when this effect is larger in magnitude than the effect of the quality difference between the two options, then inferior lock-in is always possible. Thus, although a larger majority can exert larger influence on an individual’s choice, the degree to which people are influenced by marginal majorities may be sufficient to determine whether lock-in on an inferior option is at all possible.

Majority influence has been one of the main research threads in social psychology following Solomon Asch’s seminal work \citep{asch1955opinions}. Nonetheless, the potential impact of marginal majorities has been largely overlooked as more emphasis was put on the influence of larger majorities. In a comprehensive literature review, we were able to identify only a handful of studies that explicitly discuss the effect of marginal majorities \citep{mackie1987systematic, martin2002levels,erb2006large}. The idea of a discontinuity in the degree of social influence also runs counter to models of majority influence in the social and behavioral sciences, which assume that increasing group or majority size has a gradually increasing impact on the degree of influence (e.g., \citealp{latane1981social,boyd1982cultural,tanford1984social,brock2001discrete,maccoun2012burden}). As a result, the theoretical implications of marginal majorities for lock-in have also not been explored.  

After presenting our theory, we demonstrate the marginal majority effect by reanalyzing data from three multiple-world experiments on binary choice from diverse contexts (political preferences, matters of fact, matters of taste), providing extensive evidence that a marginal majority can substantially influence people’s choices, even when this marginal majority is not emphasized by experimental design. We then show that when the marginal majority effect is larger than the effect of the quality difference between the two items, the social system can lock-in, as predicted by our theory. Moreover, our empirical analysis shows that, in practice, the converse is also true, i.e., when the marginal majority effect is either non-existent or smaller than the quality difference of the two items, then the system is self-correcting \citep{goeree2007self,van2019self}. Our results thus suggest that the marginal majority effect is a strong predictor of the occurrence of lock-in in a variety of settings. We also provide the first empirical test of predictions regarding the choice probabilities at equilibrium that follow from the existing theory of social influence for binary choice \citep{arthur1989competing}.

\section*{Theory}
\label{results}

\subsection*{Model}
\label{sectionUrnFunction}
A large number of individuals are presented one after the other with a choice between two options, $A$ and $B$. The probability $p_A$ that an individual chooses option $A$ is an increasing function of the current popularity of $A$, defined as the proportion of previous individuals that have chosen this option, and it is independent of previous individuals' choices otherwise. That is, there is some non-decreasing function $f$ such that $p_A=f(x_A)$, where $x_A=\frac{n_A}{n_A+n_B}$ is the current popularity of $A$ (\cref{figEquilibria1}a). We call this a \textit{binary choice process} and $f$ the \textit{influence curve}. By convention, we assume that $A$ is the inherently less appealing option. This is defined to mean that, if the two options were equally popular, $A$ would be no more likely than $B$ to be chosen, that is, $f(0.5)\leq 0.5$.

Our model does not assume that individuals are homogeneous with respect to preferences or in the way that they are influenced by others. Instead, $f(x_A)$ represents the probability that a \textit{randomly chosen} individual chooses option $A$, given $A$'s current popularity. It can be thought of as an aggregation of individual-level influence curves $f_i(x_A)$ which can take any form, but we are not modeling them explicitly (see Appendix C). In the special case that the $f_i(x_A)$'s are step functions that take only the values $0$ and $1$, this is a threshold model with heterogeneous thresholds \citep{granovetter1978threshold}.

\subsection*{Background and definitions}

\begin{figure}[htb!]
         \centering
         \includegraphics[width=\columnwidth]{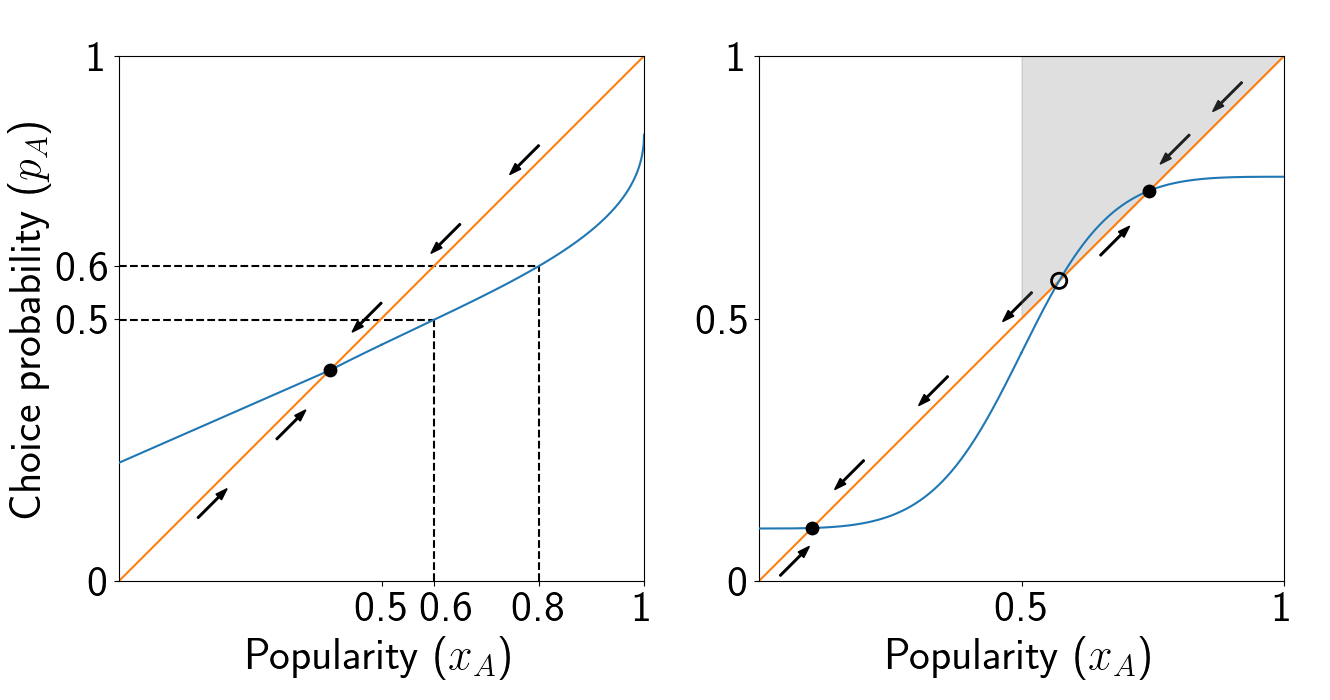}
    \caption{Influence curves and equilibria. \textbf{Left:} The probability that an individual chooses item $A$ is a function of its current popularity, that is $p_A=f(x_A)$. In the binary choice process described by this influence curve, if the popularity of item $A$ is $x_A=0.8$, then it is selected with probability $p_A=0.6<x_A$, thus driving its popularity downwards (indicated with an arrow pointing downwards along the diagonal). As the popularity decreases from 0.8 towards 0.6, the choice probability also goes down, driving the popularity further down and so on. A similar argument shows that the popularity is driven upwards whenever $p_A<x_A$. Where the graph intersects the diagonal ($p_A=x_A$), the choice probability is equal to the popularity, indicating an equilibrium. \textbf{(b)} The graph can intersect the diagonal at multiple points, implying multiple possible equilibria. At the points where the graph crosses the diagonal from above (filled circles), the equilibria are stable, meaning that deviations from those will eventually be corrected (arrows showing the direction in which popularity moves are converging). Where it crosses from below (open circle), there is an unstable equilibrium (popularity from nearby points moves away from that point). The process is lock-in-prone if and only if there exists a stable equilibrium in which the popularity of A is larger than 0.5, i.e., on the right half of the graph. This is equivalent to the curve entering the region $y>x>0.5$ (shaded region), known as the \textit{lock-in region} \citep{van2019self}.}
    \label{figEquilibria1}
\end{figure}

Because of the inherent stochasticity in individuals' choices, the popularity of the two options will initially fluctuate randomly. In particular, even if $A$ is less appealing than $B$, it is possible that $A$ will gain an initial popularity advantage, which will make it more likely to be chosen by subsequent individuals. We focus on whether it is possible for such a popularity advantage to persist in the long term \citep{arthur1989competing,van2019self}.

Under a mild technical assumption,\footnote{It is enough to assume that $f$ has at most finitely many discontinuities.} \citeauthor{hill1980strong} show that $x_A$ eventually necessarily converges to some equilibrium value \citep[Corollary 2.1]{hill1980strong}. However, the limit value to which it converges is ex ante random. Let us denote this (in principle random) limit by $x_\infty$. We have the following definition.

\begin{definition}
    A binary choice process is \textbf{self-correcting}  if the probability that a majority of individuals chooses the inherently less appealing option in the long-term is $0$, i.e., $\Pr{x_\infty>0.5}=0$. Otherwise, it is \textbf{lock-in-prone}.\footnote{This definition is related to the notion of (un-)predictability: a process is unpredictable if $x_\infty$ may take two or more values with positive probability, i.e., $\Pr{x_\infty=x_1}>0$ and $\Pr{x_\infty=x_2}>0$ for some $x_1\neq x_2$ \citep{arthur1989competing}. Lock-in-proneness is a stricter condition, requiring that one of these values corresponds to a majority of the inherently less appealing option.}
\end{definition}

Although the term ``lock-in'' sometimes is used to describe any stable majority, here we will reserve it for referring to a stable majority \textit{of the inherently less appealing option} ($x_{\infty}>0.5)$.

The graph of the influence curve $f$ can tell us whether the process is lock-in-prone or self-correcting. To avoid technicalities in the statements of the results, we will assume throughout that $f$ is non-decreasing, $0<f(x)<1$ for all $x$, $f$ crosses the diagonal line $y=x$ whenever it meets it, and $f$ does not cross the diagonal infinitely many times. The following theorem formalizes the lock-in condition in \cite{van2019self}. Related results appear in an earlier paper by \cite{arthur1989competing}. The proof is given in the Appendix.

\begin{theorem}
\label{theoremCharacterizationLockInProneness}
    A binary choice process with influence curve $f$ is lock-in-prone if and only if the graph of $f$ enters the region $y>x>0.5$.
\end{theorem}

The region $y>x>0.5$, highlighted in \cref{figEquilibria1}, is the \emph{lock-in region} defined by \cite{van2019self}. \Cref{figEquilibria1}a also gives some intuition about why the theorem holds. A special role is played by the points where the influence curve \emph{downcrosses} the diagonal line $y=x$, i.e., points $x_0$ for which $f(x)>x$ to the left of $x_0$ and $f(x)<x$ to the right. Such points are called \emph{stable equilibria} of $f$ and they are the possible long-term limits for $A$'s popularity \citep{hill1980strong,arthur1989competing}.\footnote{This is a well-known fact for deterministic dynamical systems, and in the context of social influence it has been used in the study of threshold-based models \citep{granovetter1978threshold,lopez2008social}. The model we use here is stochastic is nature. In particular, unlike the threshold model or other deterministic systems, the same exact initial conditions can lead to very different long-term outcomes. The theory developed in \cite{hill1980strong} guarantees that some of the same arguments used for deterministic systems are also valid in our case.} Because $A$ is the less appealing option ($f(0.5)\leq 0.5$), it is always possible for option $B$ to be (equally or) more popular in the long-term, hence there is always a stable equilibrium $x_0\leq 0.5$. If the influence curve enters the lock-in region (i.e., the process is lock-in-prone), then there is another stable equilibrium above $x=0.5$, implying that there are at least two possibilities for the long-term limit, making the process unpredictable \citep{arthur1989competing}.

A basic model assumption is that the choice probability depends only on the proportion of individuals that have previously chosen $A$ or $B$, no matter \textit{how many} individuals have come before. This assumption can be relaxed, and $f$ can be thought of as describing only the behavior of individuals that come \textit{late} in the sequence, with \cref{theoremCharacterizationLockInProneness} and all statements about the long-term dynamics continuing to hold \citep{arthur1986strong}.

\subsection*{Main argument: Step-like majority influence and lock-in}
\label{sectionPureRanking}

\begin{figure}
         \centering
         \includegraphics[width=\columnwidth]{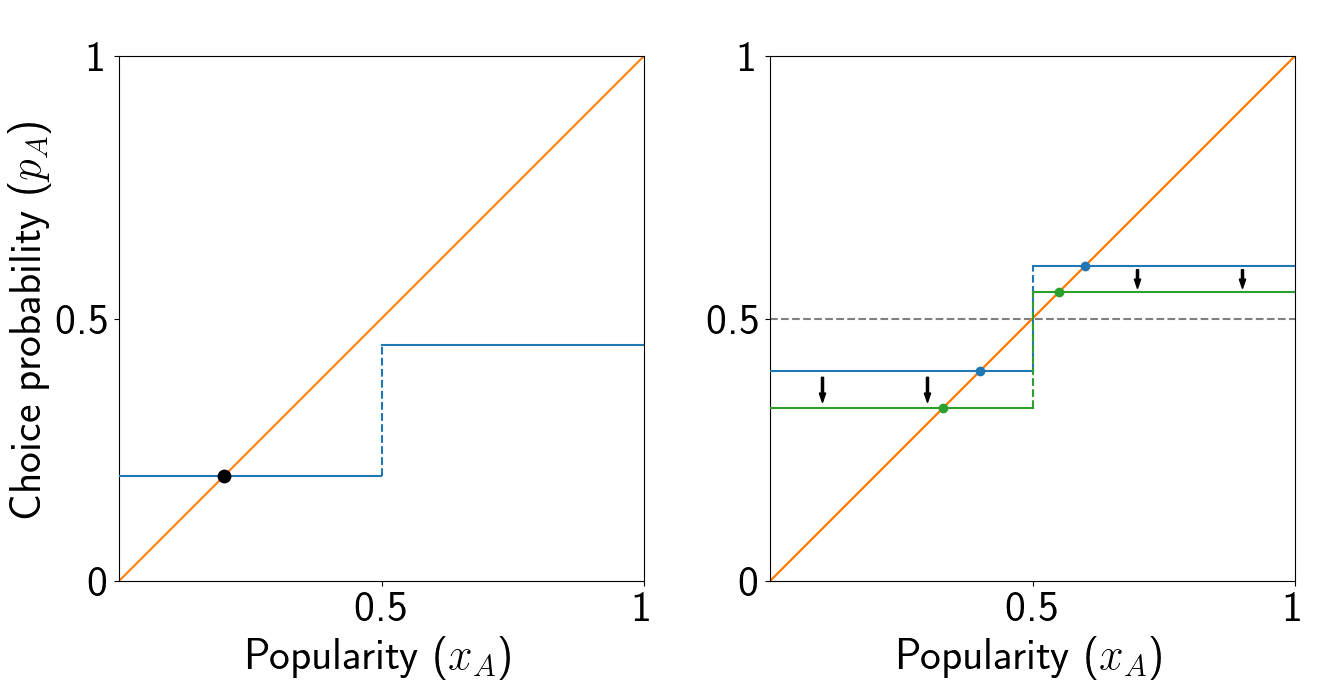}
    \centering
    \caption{\textbf{Left:} When social influence is exclusively majority-based, the graph of $f$ consists of two flat parts, on either side of $x=0.5$. \textbf{Right:} If $A$ and $B$ are of identical inherent appeal, then the graph is point-symmetric with respect to $x=y=0.5$ (blue line), with the graph entering the lock-in region immediately to the right of $x=0.5$. Substituting item $A$ with one of slightly smaller inherent appeal results in a downward shift of the two flat parts of the graph (green line). As long as the difference in inherent appeal is small, the influence curve still enters the lock-in region.}
    \label{figRanking1}
\end{figure}

Let us now suppose that individuals do not take into account the exact proportion of previous individuals who have chosen alternatives $A$ or $B$, but instead that they are informed only (or pay attention only to) whether $A$ or $B$ is currently more popular. Given that an individual does not distinguish between $A$ having popularity, e.g., $51\%$ or $95\%$, their choice probability is the same in both cases. The graph of $f$ is thus flat in each of the regions $x<0.5$ and $x>0.5$ (\cref{figRanking1}).

In this setting, for alternatives with a small difference in inherent appeal, the system is necessarily lock-in-prone. This is because a small difference in inherent appeal implies that the average of the two flat parts of the influence curve is close to $y=0.5$. As a result, because of the discontinuity at $x=0.5$, the right part of the influence curve is above $y=0.5$ and it enters the lock-in region immediately to the right of $x=0.5$ (\cref{figRanking1}b). A more general and quantitative version of the above statement is given below (\cref{theoremMultipleEquilibria}).

\subsection*{Combining step-like and continuous popularity-based influence}
\label{sectionRankingAndProportion}

We now consider a more general setting, where there can be both step-like and continuously increasing influence. This can model situations in which some individuals pay attention to which option is the more popular one, while others are subject to a gradually increasing influence as the popularity of one of the options increases. It is also possible that the same individual is subject to both types of influence. For example, one might be weakly inclined to align with the majority even if this majority is small and at the same time become more convinced that the majority is correct as the majority grows.
 
\begin{figure}
         \centering
         \includegraphics[width=1\columnwidth,trim = 0 0 0 0, clip]{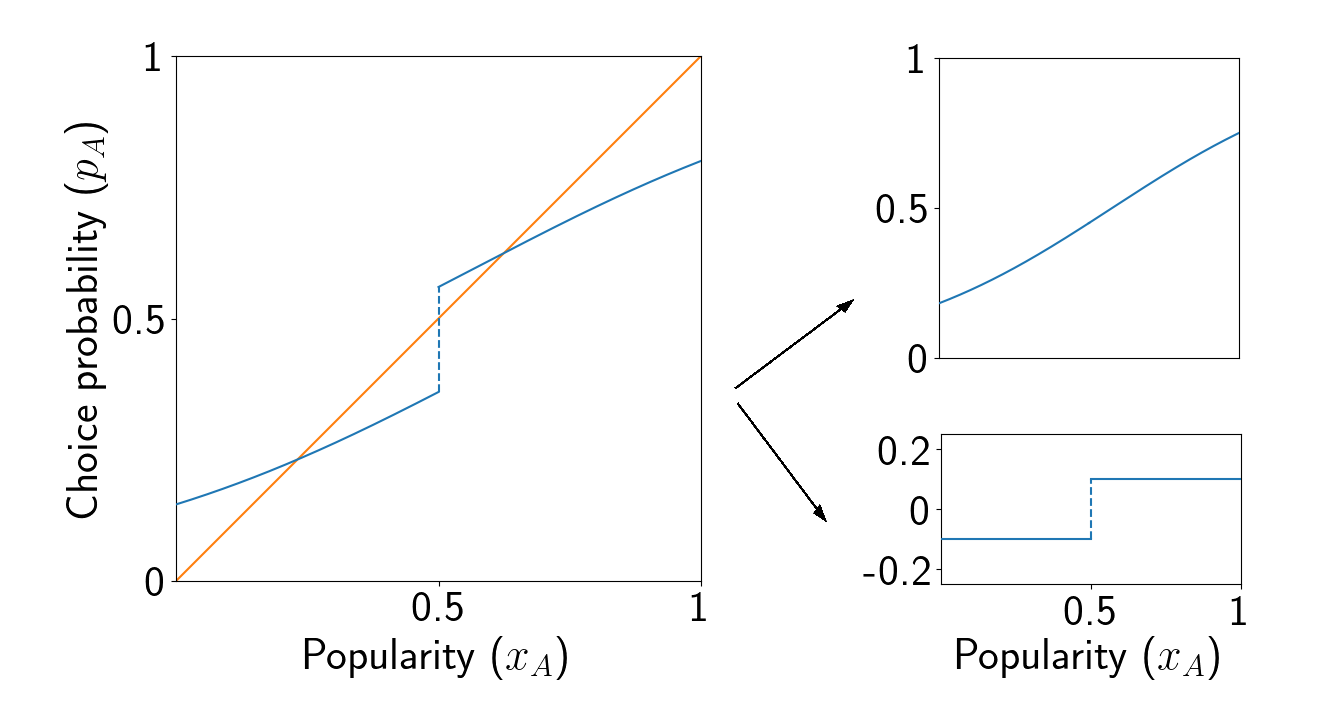}
         \centering
    \caption{Choice probability vs popularity in the presence of both majority-based and continuously increasing influence. \textbf{Left:} The marginal majority effect $M$ can be defined as a discontinuity in the influence curve at $x=0.5$. \textbf{Right:} An influence curve $f$ with a non-zero marginal majority effect can be written as the sum of two parts, $f(x)=g(x)+\frac{M}{2}\cdot u(x)$, where $g$ is continuous and $u$ is a step function, with $u(x)=\pm 1$ if $x\gtrless 0.5$.}
    \label{figRankAndProp1}
\end{figure}

In this more general setting, social influence can be modelled through an arbitrary increasing influence curve with a discontinuity at $x=0.5$ (\cref{figRankAndProp1}a). Such an influence curve can be decomposed into a continuous part and a purely majority-dependent part by writing
\begin{equation}
\label{eqDecompose1}
    f(x)=g(x)+\frac{M}{2}\cdot u(x),
\end{equation}
where $g$ is the continuous part,\footnote{If $f$ has further discontinuities, $g$ might also be discontinuous at points $x\neq 0.5$. Our results rely on the fact that $g$ is continuous at $x=0.5$, which holds by definition.} $M$ is the size of the jump at $x=0.5$, and $u(x)$ is a step function that takes the value $1$ or $-1$ depending on whether $x>0.5$ or $x<0.5$ (and $u(0.5)=0$) (\cref{figRankAndProp1}b). We call $M$ the \emph{marginal majority effect}.

As in the case of purely step-like majority influence, if the jump that occurs at $x=0.5$ is large enough to overcome the difference in inherent appeal of the two options, the graph of $f$ will enter the lock-in region immediately to the right of $x=0.5$, making the system necessarily lock-in-prone (\cref{figRankAndProp1}a). To make this precise, let us define the difference in inherent appeal $d$ as the difference in choice probability of $B$ vs $A$ when the two items are equally popular, that is when $x_A=0.5$. Note that when $x_A=0.5$, the choice probability of $A$ is $g(0.5)$ and that of $B$ is $1-g(0.5)$, hence $d=(1-g(0.5))-g(0.5)=1-2\cdot g(0.5)$.

\begin{theorem}
\label{theoremMultipleEquilibria}
If $M>d$, then the system is lock-in-prone.
\end{theorem}

This theorem provides only a sufficient condition for lock-in-proneness. However, this condition correctly predicts most cases of lock-in in several social influence experiments that we analyze (see Empirical Evidence section). In other words, for the experiments analyzed below, the condition $M>d$ turns out to be not only sufficient, but also approximately necessary.

Compared to \cref{theoremCharacterizationLockInProneness}, which is an if-and-only-if statement, checking the condition $M>d$ requires much less data, because it requires knowing the values of the influence curve only near $x=0.5$, while \cref{theoremCharacterizationLockInProneness} requires knowing the influence curve for all $x>0.5$.\footnote{If we do not know ex ante which of the two options is inherently more appealing, then we need the influence curve for all $x$.}$^,$\footnote{Since the condition of \cref{theoremCharacterizationLockInProneness} is both necessary and sufficient, in theory it is satisfied whenever the condition of \cref{theoremMultipleEquilibria} is satisfied. However, in practice, the quantities relevant for the two conditions are estimated differently, hence it is possible for the condition of \cref{theoremMultipleEquilibria} to be satisfied in a certain dataset without that of \cref{theoremCharacterizationLockInProneness} being satisfied.} \Cref{theoremMultipleEquilibria} thus provides a more parsimonious way to assess whether a system can exhibit lock-in.

\subsection*{Continuous influence curves}
\label{sectionContinuousUrnFunctions}

\Cref{theoremMultipleEquilibria} implies that a discontinuity in the influence curve is sufficient for lock-in-proneness, as long as the difference in inherent appeal of the two options is small. In contrast, a process with a continuous influence curve does not have to be lock-in-prone, even if the difference in inherent appeal is small. In \cref{figEquilibria1}, for example, even if the value of $f$ was very close to the diagonal at $x=0.5$, which would imply a tiny difference in inherent appeal, $f$ would not enter the lock-in region.

\begin{figure}
    \centering
    \includegraphics[width=\columnwidth]{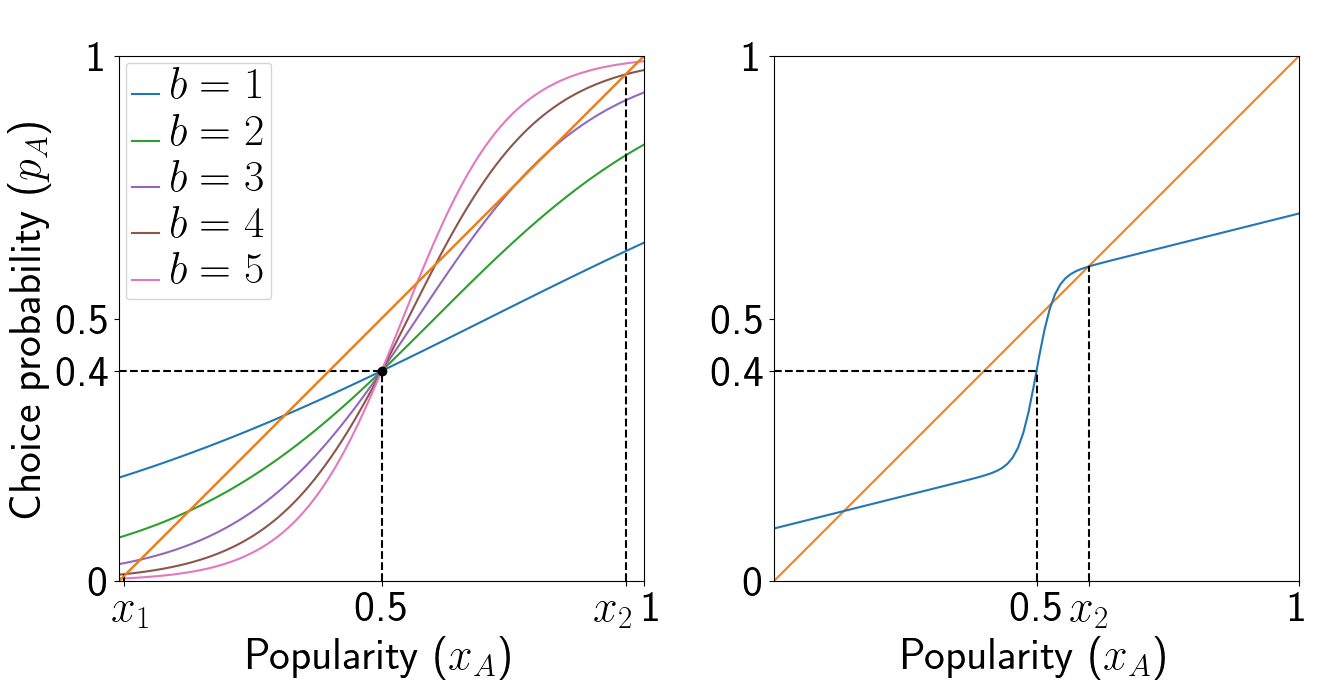}
    \caption{\textbf{Left:} Logistic influence curve (\cref{eqLogistic}) with $d=0.2$ and varying $b$. For $b<3$, the process is self-correcting, while for $b>3$ it is lock-in-prone with stable equilibria at $x_1\approx 0$ and $x_2\approx 1$ (here shown for the case $b=4$).
    \textbf{Right:} Given that $A$ is an inherently less appealing option ($f(0.5)$ is below the diagonal), between $x=0.5$ and any downcrossing on the right half of the line ($x_2$) there is an upcrossing. As a result, for a downcrossing to occur above but close to $x=0.5$, $f$ must have a very steep slope near $x=0.5$ which quickly diminishes shortly after, i.e., it must exhibit a marginal majority-like effect.}
    \label{figSensitivity}
\end{figure}

This is not to say that a binary choice process with a continuous influence curve \emph{cannot} be lock-in-prone. \Cref{figSensitivity}a shows the graph of the logistic function
\begin{equation}
\label{eqLogistic}
    f(x)=\frac{1}{1+\frac{1+d}{1-d}\cdot e^{b(1-2x)}},
\end{equation}
a standard function used to model binary choice in the social, behavioral, and management sciences, which has been often used to describe forms of social influence \citep{brock2001discrete,blume2003equilibrium,bruch2006neighborhood,van2009neighborhood,maccoun2012burden,frey2021social}. Here $d$ is the difference in inherent appeal as defined earlier (in the graph of \cref{figSensitivity}a, $d=0.2$),
while $b$ controls the slope. For small values of $b$, the graph does not enter the lock-in region, making the system self-correcting, but for larger values ($b>3$ when $d=0.2$) the graph does enter the lock-in region, making the system lock-in-prone. In the latter case, however, the two stable equilibria of the system lie very close to $x=0$ and to $x=1$, respectively, which correspond to winner-take-all situations. This is the case because such large values of $b$ imply very strong social influence.

More precisely, the logistic equation has the property that social influence is relatively much stronger when $x_A$ approaches extreme values ($0$ or $1$), compared to when it takes more moderate values. This is common in systems that exhibit network externalities \citep{katz1994systems}. When this is the case, equilibria are likely to occur only at values $x<0.5$ or \emph{well above} $x=0.5$, as suggested by fig. \ref{figSensitivity}a. In contrast, when social influence is highly sensitive to differences in popularity \emph{only} near $x=0.5$ (e.g., in the presence of marginal majority effects in otherwise relatively flat influence curves), equilibria just above $x=0.5$ are possible (figs. \ref{figRanking1}b, \ref{figRankAndProp1}a, and \ref{figSensitivity}b).

\begin{table}[htbp]
\centering
\begin{tabular}{l|c|c|c|l}
 Dataset & Publication & \#items & \#trials & \#participants  \\ 
 \hline
V2019 & \cite{van2019self} & 7 & 2 & 530 \slash\ 3500\\ 
MDRT2019 & \cite{macy2019opinion} & 20 & 8 & 220 \\ 
FV2021 & \cite{frey2021social} & 30 & 30-45 & 15 \slash\ 100
\end{tabular}
\caption{The multiple-world experiments that we re-analyze in this study, with the number of items, number of worlds (trials) per item, and number of participants per world in each experiment. V2019 had 1 world with 530 participants and 1 with 3500 participants per item. FV2021 had 10-15 worlds of 100 participants and 20-30 worlds of 15 participants for each item.
}
\label{multipleWorldDatasets}
\end{table}

\section*{Empirical evidence}

We test the predictions of our theoretical framework against the findings of experiments on binary choice under social influence. We analyze data from three recent multiple-world experiments that have reported differential findings regarding the possibility of lock-in, namely the ones by \cite{van2019self} (no lock-in), \cite{macy2019opinion} (lock-in commonly observed), and \cite{frey2021social} (lock-in observed only in some of the experimental items), to which we refer below as V2019, MDRT2019, and FV2021, respectively (\cref{multipleWorldDatasets}). These experiments are suitable for empirically testing our theoretical findings because they involve choice between two alternatives only, participants choose in a sequence, and each participant has aggregate information about the choices of all past participants in the same trial.

In the analysis below, we first estimate the influence curves and evaluate empirically the predictions regarding the possibility of lock-in and the long-term popularity that follow from previous theory \citep{van2019self,arthur1989competing} (see \cref{theoremCharacterizationLockInProneness}). We then check whether marginal majority effects are present and, if yes, whether the condition $M>d$ predicts the possibility of lock-in (\cref{theoremMultipleEquilibria}, also see Methods section for details). The analysis reported in the body of the paper uses all the data available both for estimating influence curves/marginal majority effects and for examining the occurrence of lock-in. In the SI, we report an ancillary analysis that uses separate subsets of the data for estimating the influence curves/marginal majority effects and for testing the occurrence of lock-in. The results of this out-of-sample analysis are similar to those reported here.

\begin{figure}[tbh!]
    \centering
    \textbf{\hspace{0.7cm} V2019}\par\medskip
    \includegraphics[width=0.6\columnwidth, trim = 0 0 0 60, clip]{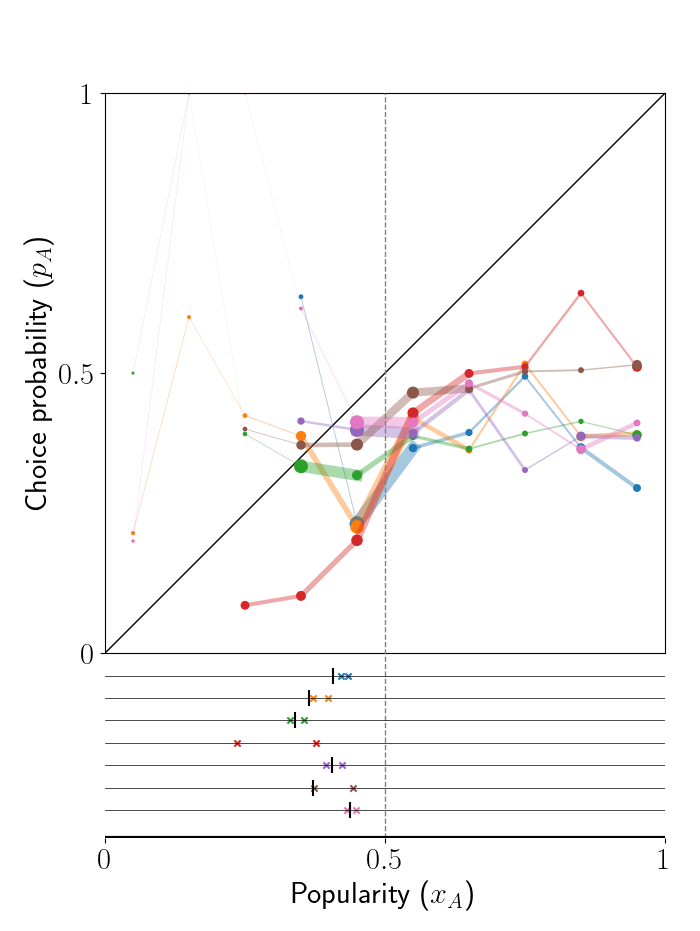}
    \caption{Influence curve estimation and end-of-trial popularities in V2019. \textbf{Top:} For each question, influence curves are estimated by grouping answers from participants in all trials into 10 intervals based on the popularity of option A at the time (see Materials and Methods). The size of each circle is proportional to the number of data points in the interval and the width of each line is proportional to the minimum number of data points in the two intervals that it connects. Given that none of the influence curves enters the lock-in region, the theory predicts that lock-in is impossible in all cases (\cref{theoremCharacterizationLockInProneness}) \textbf{Bottom:} Colored crosses indicate the end-of-trial popularities $x_{\infty}$ of option $A$ in the two trials of the experiment (one horizontal line per question). In all cases $x_{\infty}<0.5$. The points where the influence curves downcross the diagonal are indicated with vertical bars. These points are the predicted long-term proportions of $A$ and they agree reasonably with the observed end-of-trial proportions. Note that in one of the two trials for each question, option $A$ was given a large artificial advantage in popularity, hence the lack of lock-in is strongly indicative of its impossibility.}
    \label{fig:influenceCurves-vdRijt}
\end{figure}

\subsection*{Datasets}

\subsubsection*{Dataset 1 - V2019 \citep{van2019self}}

This experiment included seven binary choice questions that were primarily on matters of taste and had no correct answer (e.g., choose the piece of art that you like more). Participants were shown the number of previous participants that had chosen each of the two answers. There were two independent worlds: in the first world there were approximately 530 participants who were informed of the true number of previous answers; in the second world there were 3,500 participants, but the less attractive option (identified in a pre-study) was given an artificial initial advantage of about 110 to 10 in terms of previous answers. The true counts of answers during the experiment were added to this artificial initial count. The influence curves for the seven questions are shown in \cref{fig:influenceCurves-vdRijt} (see Materials and Methods).

\begin{figure}[tbh!]
    \centering
    \textbf{\hspace{0.7cm} MDRT2019}\par\medskip
    \includegraphics[width=0.6\columnwidth, trim =0 0 0 70, clip]{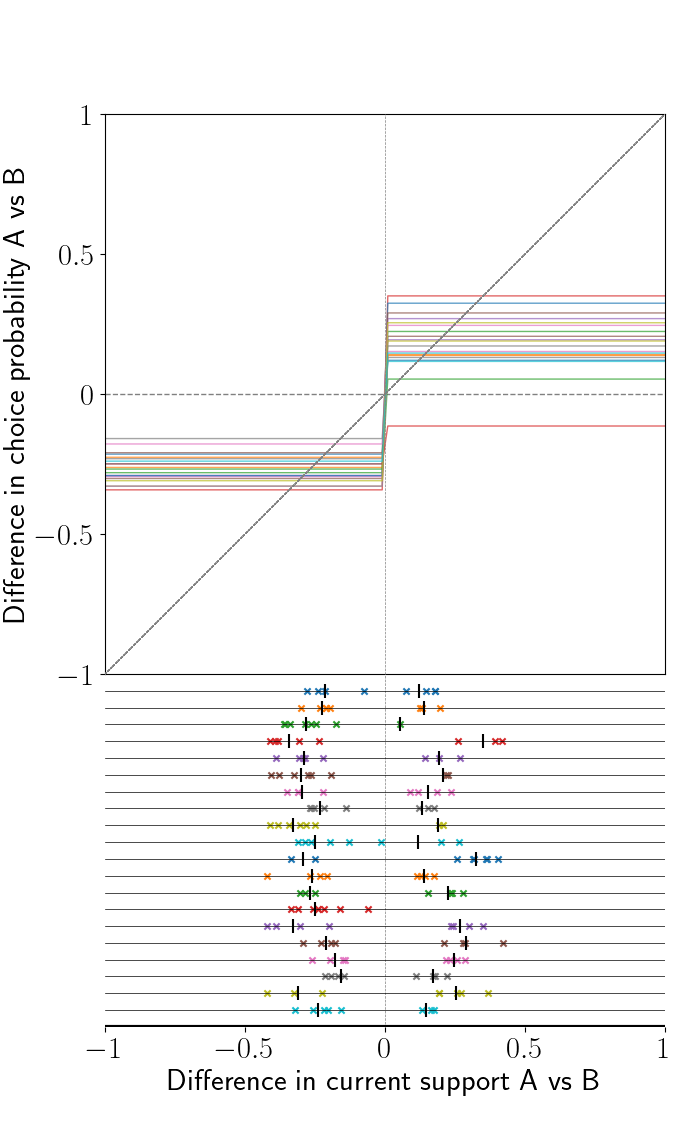}
    \caption{Influence curves for MDRT2019. Participants were informed whether the support for a statement so far was larger among democrats or republicans, hence the variables of interest are the difference in the percentage support so far between the two parties and the difference in choice probability between the two parties, both in the range $[-1,1]$. For each statement, party A is defined as the party that is inherently less likely to support the statement. The lock-in-proneness condition is that the influence curve enter the region $y>x>0$. Because participants did not know the exact difference in current support, but only whether it was positive or negative, we force the influence curves to be flat on either side of $x=0$ and make a single estimate on each side (see Materials and Methods). In all but one case the influence curve enters the lock-in region, predicting that lock-in is possible. In all but the same one case, end-of-trial majority support by party A ($x_{\infty}>0$) is observed in at least one of the trials.}
    \label{figInfluenceMacy}
\end{figure}

\subsubsection*{Dataset 2 - MDRT2019 \citep{macy2019opinion}}

Participants in the United States reported their political orientation (democrat or republican) and then answered whether they agreed or not with a series of 20 statements on social issues. For each statement, the participant was informed whether the proportion of participants supporting the statement was larger among republicans or democrats in their world (in 8 independent worlds + 2 control worlds), but not by how much. This was communicated both in a text message on the screen and by the font color of this message (red or blue). Thus, in this experiment, a marginal majority must have the same effect as a much larger majority by design, because the information provided to the participants in both cases was the same.

Furthermore, because the information provided involved the comparison of two proportions (one for democrats and one for republicans), a slight modification to our framework is needed to apply it to this case. Specifically, a larger republican (democrat) support can be sustained if, given such a larger republican (democrat) support, the probability of agreeing with the statement is also larger for republicans (democrats). The variables of interest here are the \emph{difference} in the two proportions ($x=x_A-x_B$) and the \emph{difference} in the choice probabilities ($y=p_A-p_B$) for the two groups, both taking values in $[-1,1]$ (\cref{figInfluenceMacy}). With the convention that $A$ denotes the party that is inherently less likely to support a statement, the lock-in region is now the region $y>x>0$.

\subsubsection*{Dataset 3 - FV2021 \citep{frey2021social}}

\begin{figure*}[htb!]
    \centering
    \textbf{\hspace{0.7cm} FV2021}\par\medskip
    \includegraphics[width=1\columnwidth]{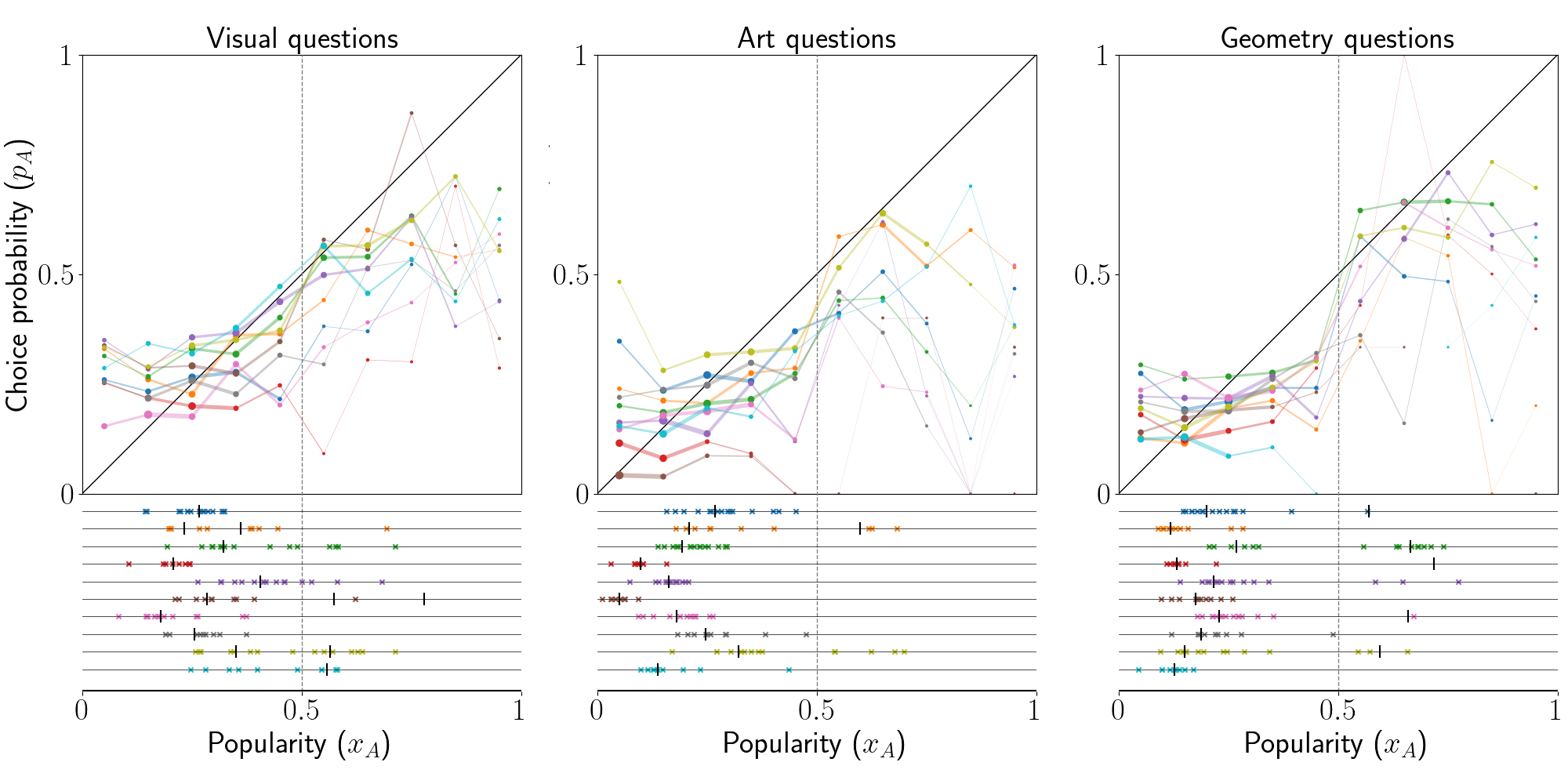}
    \caption{Similar to \cref{fig:influenceCurves-vdRijt}, but for the FV2021 dataset. In this experiment one of the two possible answers for each question was correct and participants were incentivized to answer correctly. Option $A$ is defined as the less popular option in the control condition, independently of being the correct one or not. There were 10 visual judgement questions (\textbf{left}), 10 art history questions (\textbf{middle}) and 10 geometry questions (\textbf{right}). Across panels, the lock-in region is entered in 8 cases and in 7 of them end-of-trial majorities of option $A$ are observed (in at least one of the trials), which is evidence for lock-in. There are also 5 cases in which the lock-in region is not entered, but end-of-trial majorities for $A$ are still observed (false negatives). Only end-of-trial proportions from longer trials (100 participants) are taken into account and shown in the bottom panels.}
    \label{fig:Frey-vdRijt}
\end{figure*}

Participants were given 30 trivia tasks, 10 of which were visual puzzles, 10 were art-related questions and 10 were geometry questions, and they were informed about the number of previous participants in their world that had chosen each of the two possible answers. For each question, there were either 10 or 15 worlds with 100 participants each and either 20 or 30 worlds with 15 participants each. Participants were allowed to abstain. Each question in this study had a definite correct answer. However, for consistency with the other datasets and the theory, we define option A (the inherently less appealing option) as the option that fewer people selected in the control condition (no social influence), independently of whether it was the correct answer or not. The influence curves are estimated as before and shown in \cref{fig:Frey-vdRijt}.

\subsection*{Results}

\subsubsection*{Influence curves predict lock-in}
We will say that ``lock-in occurs'' if the end-of-trial choice proportions favor the less appealing option (i.e., option A) in at least one of the trials. 

In V2019, the influence curves do not enter the lock-in region for any of the 7 questions (\cref{fig:influenceCurves-vdRijt}) and, as predicted, no lock-in occurs. 
Although in this experiment there were only two trials per question, in one of these trials option A was given a large initial advantage, hence the absence of lock-in is strongly indicative of its impossibility. In MDRT2019, the influence curves enter the lock-in region in 19 out of 20 cases and lock-in occurs in the same 19 questions ($0$ misclassified items). Finally, in FV2021, the influence curves enter the lock-in region in 8 cases, in 7 of which lock-in occurs. Lock-in also occurs in 5/22 of the items for which the influence curve does not enter the lock-in region, for a total of 6 misclassified items. The results are similar in an out-of-sample prediction analysis (see SI).

The misclassifications obtained for some questions can be attributed to several reasons. First, we define lock-in based on the end-of-trial proportions. It is possible that trajectories in which option A had an end-of-trial majority would have eventually been corrected had the trial been longer. This is especially likely for questions in which the end-of-trial proportions have high dispersion (see for example items 5 and 9 in \cref{fig:Frey-vdRijt}a), implying lack of convergence.\footnote{The opposite side of this is that, for some of the questions for which lock-in is predicted, it is possible that they are not observed in any of the trials of the experiment because they occur rarely. However, in the only case where lock-in was predicted and not observed, the prediction was highly unreliable (item 4 in \cref{fig:Frey-vdRijt}c).}

Another source of error is of course estimation error; for example, for a few questions in FV2021, the influence curves get very close to the lock-in region without entering it. Apart from the inherent noise in estimating the influence curve with a finite amount of data, note that we are also averaging over bins of size 0.1; it is likely that the true influence curve enters the lock-in region in a subset of the bin. However, choosing a smaller bin size would have reduced the data per bin and increased the noise.

\subsubsection*{End-of-trial proportion prediction}

An important prediction of the theoretical model we have used is that the stable equilibria of the influence curve (points where it downcrosses the diagonal) are the possible long-term limits for the popularity \citep{arthur1989competing}. This prediction is visually confirmed in  \cref{fig:influenceCurves-vdRijt,fig:Frey-vdRijt,figInfluenceMacy}: The points in the bottom panel are positioned below the stable equilibria suggested by the influence curves in the top panels. The results are similar in an out-of-sample prediction analysis (see SI).

\subsubsection*{Marginal majority effects}

The marginal majority effects are most straightforward in MDRT2019, where social influence is purely majority-driven. The discontinuities of the influence curves at $x=0$ in this experiment vary between $0.13$ and $0.7$. In V2019, all influence curves are relatively flat away from the middle, while for several of the questions we observe a sharp increase in the influence curve between the two central bins (changes in the range $0.07$-$0.23$ for five of the questions), which is indicative of a marginal majority effect. In other words, most of the social influence observed is a result of a marginal majority or marginal majority-like effect. Note that in this experiment the majority choice was not made visually salient on the experiment's screen (the positions of the options were fixed), but it was instead inferred by the participants from the number of choices indicated under the items. Despite the marginal majority effect being substantial in several cases, it was not large enough for the influence curve to enter the lock-in region in any of the questions. The reason is that in cases in which the marginal majority effect was large, the difference in inherent appeal of the two options happened to also be large.

In FV2021 there is also a sharp increase in choice probability as popularity crosses $x=0.5$, while the choice probability is relatively insensitive to popularity away from $x=0.5$ (\cref{fig:Frey-vdRijt}). This effect is especially pronounced for the geometry questions and to a somewhat lesser extent for the art questions. The size of the change between the two central bins varied substantially  across questions, but it exceeded $0.15$ in 8 out of 27 questions for which there was data. In many cases this increase was large enough for the influence curve to enter the lock-in region immediately to the right of $x=0.5$.

\subsubsection*{Using marginal majority effects and inherent appeal differences to predict lock-in}

\begin{figure}[tbh!]
    \centering
    \includegraphics[width=0.6\columnwidth,trim = 0 0 0 40, clip]{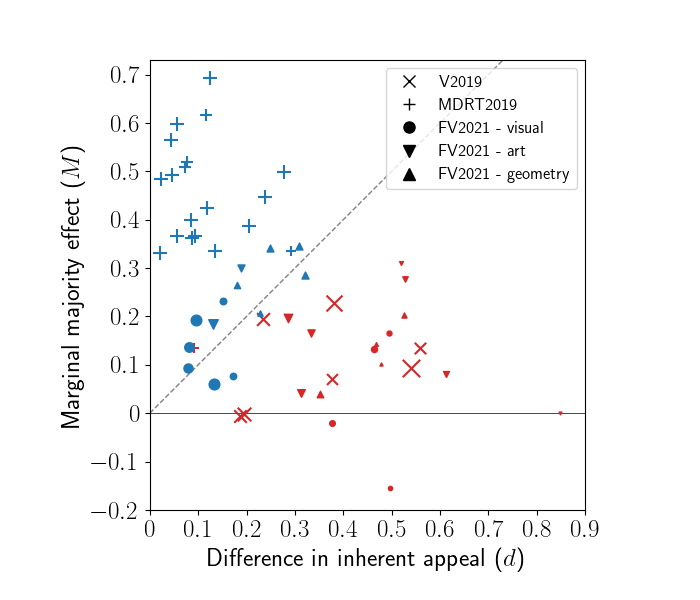}
    \caption{Summary of marginal majority effect vs inherent appeal difference, for all datasets. Each data point corresponds to one experimental item (question). The marginal majority effect $M$ is estimated as the increase in the influence curve between the two central bins (see \cref{fig:influenceCurves-vdRijt}). The difference in inherent appeal $d$ is estimated as the difference in choice probability of the two options in the control experiments (no social influence), with suitable modifications in the case of MDRT2019, where $d$ is a measure of the inherent ideological content of an opinion (see Methods). Blue points correspond to questions in which lock-in was observed, i.e., option $A$ had an end-of-trial majority in at least one of the trials, while red color indicates the absence of lock-in. The size of the data points is proportional to the data available for the estimation of $M$. The theory predicts that for all points above the dashed gray line ($M=d$) lock-in is possible, while no prediction is made for points below the dashed line. It turns out that for the datasets examined the condition $M>d$ is highly accurate even as an if-and-only-if condition for the possibility of lock-in.}
    \label{figSummary}
\end{figure}

\Cref{figSummary} summarizes the data from the three experiments and shows that the condition identified in \cref{theoremMultipleEquilibria}, namely that the marginal majority effect exceed the difference in inherent appeal of the two options ($M>d$), is a good predictor of lock-in-proneness.\footnote{The definitions of $M$ and $d$ are suitably modified in the case of MDRT2019, to reflect the fact that the influence curve is the difference of two choice probabilities (see Methods). Theorem 2 remains valid in this case (see Appendix).} Specifically, in almost every case that this condition is satisfied, lock-in is observed (27/28 cases, or 96\%), i.e., the inherently less appealing option has an end-of-trial majority in at least one of the trials. And although \cref{theoremMultipleEquilibria} makes no statement when $M\leq d$, lock-in rarely occurs in this situation (5/29 cases, or 17\%). In other words, while \cref{theoremMultipleEquilibria} identifies $M>d$ as merely a sufficient condition for lock-in-proneness, in practice it works as both sufficient and necessary,\footnote{The proof of \cref{theoremMultipleEquilibria} suggests that this should be the case if the most likely place for the influence curve to enter the lock-in region is immediately to the right of $x=0.5$.} being equally accurate with the true if-and-only-if condition of \cref{theoremCharacterizationLockInProneness}, despite requiring much less data to be checked.

\Cref{figSummary} also suggests an explanation for the discrepancy in terms of the possibility of lock-in across experiments: in MDRT2019, the marginal majority effect was substantial in almost all of the cases and the difference in inherent appeal was moderate, which explains why lock-in was observed in all questions but one. In V2019, the majority effect was substantial only in about half of the questions, but for those questions the two options happened to differ a lot in terms of inherent appeal. As a result, lock-in was not possible. Had the two options been of comparable inherent appeal, lock-in would likely have occurred. Finally, in FV2021, both the differences in inherent appeal and the marginal majority effects varied in magnitude, leading to lock-in only in a subset of the questions.

\section*{Discussion}

Majority influence has been one of the main research topics in the study of social influence in psychology and beyond \citep{asch1955opinions,levine2012majority,kruglanski1990majority}. The current conception of majority influence, as it has been also consolidated in formal models in the social and behavioral sciences, assumes that social influence gradually increases as the size of the majority increases \citep{latane1981social,tanford1984social,brock2001discrete,maccoun2012burden}. In this work, we defined and empirically demonstrated the marginal majority effect, a discontinuous increase in the choice probability of an option when it surpasses a competing option in popularity, and showed that it is a powerful mechanism of influence operating above and beyond the previously purported gradually increasing majority influence. In addition, we showed theoretically that when a large number of people choose from a given pair of options, the marginal majority effect provides a mechanism for lock-in. More precisely, we showed that lock-in may occur when the marginal majority effect exceeds the difference in inherent appeal of the two options (\cref{theoremMultipleEquilibria}). Finally, we estimated the marginal majority effects and the differences in inherent appeal between options directly from data in three recent sequential choice experiments with several cases each, and showed that our model accurately predicts the presence of lock-in in almost all cases, resolving a recent conundrum in the empirical literature.

The marginal majority effect leads to a qualitatively different type of lock-in than previously identified. Past theoretical studies of lock-in have generally focused on systems that exhibit strong network externalities \citep{katz1985network}, such as in the context of competing technologies \citep{david1985clio,arthur1994increasing}, coordination problems \citep{hodgson2004complex}, and the formation of social norms \citep{young1993evolution}. In the presence of network externalities, the value of choosing the more popular option is ever increasing with its popularity, or equivalently, the less popular option becomes increasingly less attractive as its popularity drops. The outcome is a spiraling feedback loop, reflected in the commonly used terms \emph{rich-get-richer dynamics} or \emph{cumulative advantage} which are often associated with winner-take-all outcomes, i.e., with virtually everyone adopting the technology or social norm that wins the competition. In contrast, majorities of small magnitude (e.g., between $50\%-60\%$) are often thought to be unstable or transient, and are theoretically impossible in many well-established models of sequential choice and cumulative advantage \citep{banerjee1992simple,bikhchandani1992theory,arthur1989competing}. Here we have shown theoretically and observed empirically that in the presence of marginal majority effects, small winning margins can be stable and become increasingly difficult to overturn over time, without necessarily leading to winner-take-all outcomes. We also demonstrated that this cannot occur for more classical models for which social influence is a gradually increasing function of popularity.

Our theoretical and empirical demonstration of the marginal majority effect and its relation to lock-in relied on the concept of influence curves, a natural framework to study binary choice under social influence. The first use of influence curves dates back to \cite{simon1954bandwagon}, who employed them as a tool to understand the limits in the predictability of election polls. In the dynamic (sequential) stochastic choice setting, influence curves have been used by \cite{boyd1982cultural,boyd1988culture} to describe the inheritance of cultural traits and by \cite{arthur1989competing} to model technology adoption. Similar frameworks, but with deterministic dynamics, are presented in \cite{schelling1978micromotives}, \cite{granovetter1978threshold}, and  \cite{lopez2008social}. The deterministic  case can be seen as a special case of the stochastic framework, where the influence curve  of any given individual is a binary-valued function (see also Appendix C). 

Within this framework, our work stands out in relating behavioral properties to the possibility of lock-in. Some early results in this direction appear in Granovetter’s seminal paper on threshold models \citep{granovetter1978threshold}, where he shows that a narrow distribution of individual adoption thresholds can prevent a behavior from spreading among people, thus locking the population in on non-adoption. \cite{arthur1989competing}, \cite{lopez2008social}, and \cite{van2019self} make the distinction between systems with one or two stable equilibria, but they do not explore the behavioral properties that may put a system into one regime or the other. Other modeling traditions, like the Bass diffusion model \citep{bass1969new}, do not generally implement bi-stable dynamics, focusing on the speed that an innovation spreads rather than whether it will spread at all. Finally, classic decision-theoretic models, such as the information cascade and herding model \citep{banerjee1992simple,bikhchandani1992theory}, are not suitable for studying the question of when lock-in is likely, because in these models lock-in is always possible by construction.

Other scholars have previously reconstructed influence curves from classic experiments \citep{efferson2008conformists} or new experimental data \citep{efferson2008conformists,morgan2012evolutionary,morgan2015development,danchin2018cultural}. Here we go a step further and use the reconstructed influence curves to test theoretical predictions regarding the long-term behavior of a system, both those following from our theory and some that follow from previous theory, including the prediction that the stable equilibria of an influence curve are the possible long-term values for popularity \citep{hill1980strong,arthur1989competing}. We find evidence for the accuracy of the predictions of both our results regarding the possibility of lock-in and those of older theory (see also SI).

As mentioned earlier, the role of marginal majorities in driving social influence has been largely ignored. In one of the experiments we analyzed (MDRT2019), participants were given information only about which of two popularity percentages was larger, not by how much, thus marginal majority effects were induced by design. As the authors of the original study have pointed out \citep{macy2019opinion}, providing only such binary information, rather than quantitative popularity information, can make it harder for a majority to be overturned, because it prevents the progressive weakening of the popularity signal when participants choose the less popular option (until a large number of participants have made the same choice and the majority option changes). Our analysis here provides theoretical support for this observation. We further note that in such binary choice settings, in which the choice probability is only affected by which option is the majority choice, it is possible to derive even quantitative predictions about the probability of lock-in \citep{analytis2022sequential}.

In contrast to MDRT2019, in the V2019 and FV2021 datasets the information about what the majority of previous participants had chosen was not emphasized \citep{van2019self,frey2021social}. Even in these experiments, our analysis demonstrated the existence of sizable marginal majority effects for most of the questions, suggesting that marginal majorities might be psychologically salient even when they are not visually emphasized (see also \citealp{mackie1987systematic,martin2002levels,erb2006large}). One possible mechanism for this is if people use majorities as a simple social learning strategy when they feel uncertain about the correct answer or haven't formed an opinion yet. Indeed, there is evidence that people (as well as other animals) use a ``follow the majority’’ heuristic to guide their own decisions \citep{zhang2006majority,gigerenzer2011heuristic,laland2004social,danchin2018cultural}. When even a fraction of individuals use such a heuristic strategy, a marginal majority effect is produced at the aggregate level (Appendix C).

Further, in the FV2021 dataset, we found the marginal majority effect to be especially pronounced in the geometry questions, which were factual questions that required specific expertise, and secondarily so in the art trivia questions. The marginal majority effect was relatively smaller in the visual tasks questions, where prior knowledge was not needed. This is consistent with previous work that finds that people are more likely to copy others or to conform when they have low confidence in their own judgment \citep{baron1996forgotten,morgan2012evolutionary,moussaid2014simple,toyokawa2019social,kendal2018social}. Finally, the marginal majority effects appear to be less pronounced on questions related to matters of taste, as those in V2019, where people were asked which of two artifacts (e.g., paintings, screensavers, bowls) they preferred. This suggests that for matters of taste the condition we identified will be only satisfied when the quality differences between the options are relatively small, and that additional mechanisms might be needed in order to produce lock-in. 

In the present study we have focused on the case of two competing options. A generalization to multiple options should take into account relative majority (plurality) effects, but also more general \emph{ranking effects}, i.e., discontinuous changes in choice probabilities as an option exceeds in popularity another option in a ranking. In fact, the marginal majority effect can be seen as a special case of a ranking effect, specifically when there are only two options available. Ranking effects are especially pronounced by design in many online interfaces such as search engines, recommender systems, and online marketplaces, where items are often displayed in a list ordered by their popularity; the order that items are displayed is well-known to heavily affect users’ click behavior and choice probabilities \citep{joachims2005accurately}. These findings have been also incorporated in formal choice models that rely on a variety of behavioral assumptions to accurately predict people’s choices \citep{krumme2012quantifying,chuklin2015click,analytis2022sequential}. The well-known Music Lab experiment \citep{salganik2006experimental} was also an early demonstration that ranking items by popularity can affect user behavior and the long term popularity dynamics---in one of the experimental conditions, the songs were ranked by popularity for the users and this had an impact on the choice probabilities and overall outcomes. 

The study of popularity dynamics in choice among multiple options is becoming more relevant as people’s choices are increasingly immersed in environments that rank the options for them according to popularity or a proxy thereof. A general mathematical theory for \textit{purely} ranking-based popularity dynamics has been developed by \cite{analytis2023ranking}, providing a unified framework for models in which choice probabilities depend only on the popularity ranking \citep{ciampaglia2018algorithmic,germano2019few,gaeta2023reconciling}. Nevertheless, no such theory currently exists that allows for both discontinuous and continuous popularity-based influence, as the multidimensional analogue of the influence curve is less well understood theoretically, and the existing results do not apply to influence functions with discontinuities \citep{arthur1986strong,benaim1998recursive}. A number of computational models have been used as a remedy, often accompanied with case-specific theoretical results \citep{krumme2012quantifying, van2016aligning, abeliuk2017taming, maldonado2018popularity,berbeglia2021market}. A multi-dimensional ``influence function theory'' would allow us to study popularity dynamics under social influence in a unified way, as well as to adjudicate the existence of lock-in in multi-alternative experiments with both discontinuous and continuous popularity-based influence such as in the Music Lab. We therefore consider it a promising avenue for future research.

\subsection*{Code availability} The Python code used to perform all analyses is available at \url{https://osf.io/evzwk/}.

\appendix

\section*{Appendix A: Methods}
\label{sectionMethods}

\subsection*{Empirical influence curves}
V2019 \& FV2021: For each question, we identify the inherently less appealing option (denoted as option A) as the option that was least popular in the control condition (no social influence). To estimate the influence curve, we first associate with each participant the ``current popularity'' of option A, that is the proportion of previous participants in the participant's world (trial) that gave this answer. We then pool together participants from all worlds and separate them according to current popularity of A in bins of size $0.1$. For each bin, we find the percentage of participants who chose A. The result is the value of the influence curve in that bin. Participants that fall on the bin endpoints are counted with weight $1/2$ in each bin. In \cref{fig:influenceCurves-vdRijt,fig:Frey-vdRijt}, participants that fall between the two central bins ($x=0.5$) are excluded, in order to highlight the marginal majority effect (definition of marginal majority effect does not include $x=0.5$). See Figs. S1 and S2 for the corresponding graphs with all participants. For V2019, apart from the trials reported in the main text, we use additional data from a trial reported in the Appendix of \cite{van2019self}, in which artificial counts were added between subsequent participants. Because this data was subject to exogenous manipulations, it is only used in estimating the influence curves, not in identifying lock-ins (see below). 

MDRT2019: For each question, we identify as party A the party whose supporters were least likely to support the statement in the control condition. For each participant, we calculate the ``current party A support'' ($x_A$) and ``current party B support'' ($x_B$) separately, as the proportion of previous participants of the corresponding party that supported the statement in the participant's world, and we define $x=x_A-x_B$. The responses of participants for which $x=0$ are ignored when estimating the influence curves. The rest of the participants are grouped based on whether $x>0$ or $x<0$. The value of the influence curve for each group is estimated as $y=y_A-y_B$, where $y_A$ ($y_B$) is the percentage support of the statement among party A (B) supporters in that group.

\subsection*{End-of-trial proportions and lock-in}
V2019 \& FV2021: For each question and each trial, the end-of-trial popularity of A ($\bar y$) is the proportion of participants in the trial that answered A. We say that there is a lock-in if $\bar y>0.5$ in at least one of the trials. MDRT2019: For each statement and each trial, we calculate the end-of-trial difference in support $\bar y=\bar y_A-\bar y_B$, where $\bar y_A$ ($\bar y_B$) is the proportion of party A (party B) supporters in the trial that supported the statement. We say that a lock-in occurs if $\bar y>0$ in at least one of the trials.

\subsection*{Estimating marginal majority effects and difference in inherent appeal}
V2019 \& FV2021: We estimate the marginal majority effect as the increase in the value of the influence curve between the two central bins, i.e., from bin $(0.4, 0.5)$ to $(0.5, 0.6)$. The inherent appeal difference $d$ is estimated as $p^B_{ind}-p^A_{ind}$, where $p^A_{ind}$ ($p^B_{ind}$) is the choice probability for $A$ ($B$) in the control condition (no social influence).

MDRT2019: We estimate the marginal majority effect as the sum $M=M_A+M_B$, where $M_A$ ($M_B$) is the marginal majority effect for supporters of party A (B), i.e., the difference in the  probability of supporting a statement when this statement has larger percentage support by members of their party vs when it has larger support by members of the opposite party. The ideological content $d$ of a statement is estimated as $d=2\cdot (p^B_{ind}-p^A_{ind})$, where $p^A_{ind}$ ($p^B_{ind}$) is the support probability for the statement among party A (B) supporters in the control condition.

The coefficient $2$ in the estimate for $d$ is justified as follows: Let $f_A,f_B$ denote the influence curves for supporters of party $A$ and $B$, respectively, and write
\begin{align*}
    f_A(x) & =g_A(x)+\frac{M_A}{2}\cdot u(x)\\
    f_B(x) & =g_B(x)-\frac{M_B}{2}\cdot u(x)
\end{align*}
where $g_A,g_B$ are continuous at $x=0$, $M_A,M_B$ are as above, and $u(x)=\pm 1$ depending on whether $x\gtrless 0$ (see \cref{eqDecompose1}). The total influence curve $f(x)=f_A(x)-f_B(x)$ can thus be written as
\begin{equation}
    \label{eqDefInfluenceCurveExp2}
    f(x)=g(x)+\frac{M}{2}\cdot u(x),
\end{equation}
where $g(x)=g_A(x)-g_B(x)$ and $M=M_A+M_B$. \Cref{theoremMultipleEquilibria} continues to hold if we take $d=d_A-d_B$, where $d_A=1-2\cdot g_A(0)$ is the difference between the probability of supporting and not supporting a statement for supporters of party A when the current percentage supports by the two parties are equal, and similarly for $d_B$ (see last sentence before \cref{theoremMultipleEquilibria}). A natural estimate for $g_A(0)$ is $p_{ind}^A$ and similarly for $g_B(0)$. Since
\begin{align}
    d=d_A-d_B & =[1-2\cdot g_A(0)]-[1-2\cdot g_B(0)]\\
    & =2\cdot [g_B(0)-g_A(0)],
\end{align}
we may estimate $d$ by $2\cdot (p_{ind}^B-p_{ind}^A)$.

\section*{Appendix B: Proofs of \cref{theoremCharacterizationLockInProneness,theoremMultipleEquilibria}}

Here we prove \cref{theoremCharacterizationLockInProneness,theoremMultipleEquilibria}. In what follows $I$ and $J$ denote intervals of real numbers (we may take $I=J=[0,1]$). We use the definition of crossing/downcrossing the diagonal from \cite{hill1980strong}. The assumptions on $f$ stated before \cref{theoremCharacterizationLockInProneness} should be understood with respect to this definition.

\begin{definition}
Let $f:I\to J$ be any map. We say that $f$ \emph{crosses} the diagonal at $x_0\in I$ if for every $\epsilon >0$, there exist points $x_1,x_2\in (x_0-\epsilon,x_0+\epsilon)$ such that $f(x_1)>x_1$ and $f(x_2)<x_2$.

We say that $f$ \textit{downcrosses the diagonal} at $x_0\in I$, if there exists some $\epsilon >0$ such that $f(x)>x$ for all $x\in (x_0-\epsilon, x_0)$ and $f(x)<x$ for all $x\in (x_0, x_0+\epsilon)$.
\end{definition}

Before we proceed to the proofs of our theorems, we need a couple of lemmas.

\begin{lemma}
\label{lemmaCrossing}
Let $f:I\to J$ be an increasing function and let $a,b\in I$, $a<b$, be such that $f(a)>a$ and $f(b)<b$. Then, there exists some $c\in (a,b)$ such that $f(c)=c$.
\end{lemma}

If $f$ was assumed to be continuous, then the above statement would follow immediately from the Intermediate Value Theorem. Here we are not assuming continuity for $f$.

\begin{proof}
Let $c=\inf\{x>a:f(x)<x\}$. The assumption $f(b)<b$ implies that $c\leq b$, while by definition we have $c\geq a$. We claim that $f(c)=c$, which will also imply that $c\neq a,b$, hence $c\in (a,b)$.

Suppose  on the contrary that $f(c)\neq c$. If $f(c)<c$, then $x=\frac{c+f(c)}{2}$ has the property that $x<c$ and, since $f$ is increasing, $f(x)\leq f(c)<\frac{c+f(c)}{2}=x$, which contradicts the definition of $c$. If $f(c)>c$, then for any $x\in \left(c,f(c)\right)$ we have $f(x)\geq f(c)>x$, again contradicting the fact that $c$ is the greatest lower bound of points with the property that $f(x)<x$. We conclude that $f(c)=c$, which completes the proof.
\end{proof}

\begin{lemma}
\label{lemmaDowncrossing}
Let $f:I\to J$ be an increasing function, and suppose that it meets the diagonal at most finitely many times. Let $a,b\in I$ be such that $a<b$, $f(a)>a$, and $f(b)<b$. Then, $f$ downcrosses the diagonal at some $x\in (a,b)$.
\end{lemma}

\begin{proof}
By \cref{lemmaCrossing}, $f(x)=x$ has at least one solution, and by assumption it has finitely many solutions. Let $x_1<\ldots <x_k$ be these solutions. The quantity $f(x)-x$ cannot change sign in any interval of the form $(x_i,x_{i+1})$, because then \cref{lemmaCrossing} would imply that there is an extra solution of $f(x)=x$ in that interval. For similar reasons, $f(x)-x$ is positive on $(a,x_1)$ and negative on $(x_k,b)$. Therefore, $f(x)-x$ has constant sign on each of the intervals $(a,x_1),(x_1,x_2),\ldots ,(x_{k-1},x_k),(x_k,b)$. Because this sign is positive on $(a,x_1)$ and negative on $(x_k,b)$, we conclude that there exists some $i$ such that $f(x)-x$ is positive on $(x_{i-1},x_i)$ and negative on $(x_i,x_{i+1})$ (where we define $x_0=a$ and $x_{k+1}=b$), making $x_i$ a downcrossing.
\end{proof}

Apart from the above lemmas, the proof of \cref{theoremCharacterizationLockInProneness} relies on two results from \cite{hill1980strong}: i) the process may only converge to a point where $f$ crosses the diagonal (Proposition 3.1), and ii) for any point where $f$ \textit{downcrosses} the diagonal, there is positive probability that the process converges to it (Theorem 4.2).

\begin{proof}[Proof of \cref{theoremCharacterizationLockInProneness}]
    Suppose first that $f$ enters the lock-in region, that is there is some $a>0.5$ such that $f(a)>a$. By our assumptions on $f$ we have $f(1)<1$ (see paragraph before statement of \cref{theoremCharacterizationLockInProneness}). Hence, by \cref{lemmaDowncrossing}, $f$ downcrosses the diagonal at some $x^*\in (a,1)\subset (0.5,1)$. By Theorem 4.2 in \cite{hill1980strong}, $\Pr{x\to x^*}>0$.

    For the converse, let $S=\{x\in [0,1]:f(x)<x\}$ and denote by $S^o$ its interior. By Proposition 3.1b in \cite{hill1980strong}, we have $\Pr{x_{\infty}\in S^o}=0$. If the process is lock-in-prone, i.e., $\Pr{x_{\infty}>0.5}>0$, then $(0.5,1]\not\subset S^o$, hence there exists some $x^{*}>0.5$ such that $f(x^{*})\geq x^{*}$. If $f(x^*)>x^*$, this concludes the proof. If $f(x^*)=x^*$, then by our assumption that $f$ crosses the diagonal whenever it meets it, $f$ must enter the lock-in region either directly below or directly above $x^*$ (more precisely, arbitrarily close to $x^*$), again concluding the proof.
\end{proof}

\begin{proof}[Proof of \cref{theoremMultipleEquilibria}]
The right-sided limit of $f$ at $x=0.5$ is
\begin{equation}
\begin{aligned}
    c=\limd{x\to 0.5^+}f(x)=\limd{x\to 0.5^+}\left[g(x)+\frac{M}{2}\cdot u(x)\right]=g(0.5)+\frac{M}{2},
\end{aligned}
\end{equation}
because $g$ is continuous at $x=0.5$ and $u(x)=1$ for any $x>0.5$. The assumption $M>d=1-2\cdot g(0.5)$ can be written as $M/2+g(0.5)>1/2$, hence we get $c>1/2$. Since $f$ is increasing, for any $x\in (0.5,c)$ we have $f(x)\geq \limd{x\to 0.5^+}f(x)=c>x$, hence $f$ enters the lock-in region. The result now follows from \cref{theoremCharacterizationLockInProneness}.
\end{proof}

The proof in the case that $f$ is given through \cref{eqDefInfluenceCurveExp2}, with the definitions of $g(x)$, $M$, $u(x)$, and $d$ that follow it, is completely analogous. Note that an analogue of \cref{theoremCharacterizationLockInProneness} is also required, which in the case where $f$ is constant on $x>0$ and on $x<0$ follows by Corollary 2.11 in \cite{analytis2023ranking}.

\subsection*{Appendix C: From individual to aggregate influence curves}
\label{appendixIndividualInfluenceCurves}

\begin{figure}[tbh!]
    \centering
    \includegraphics[height=0.47\columnwidth]{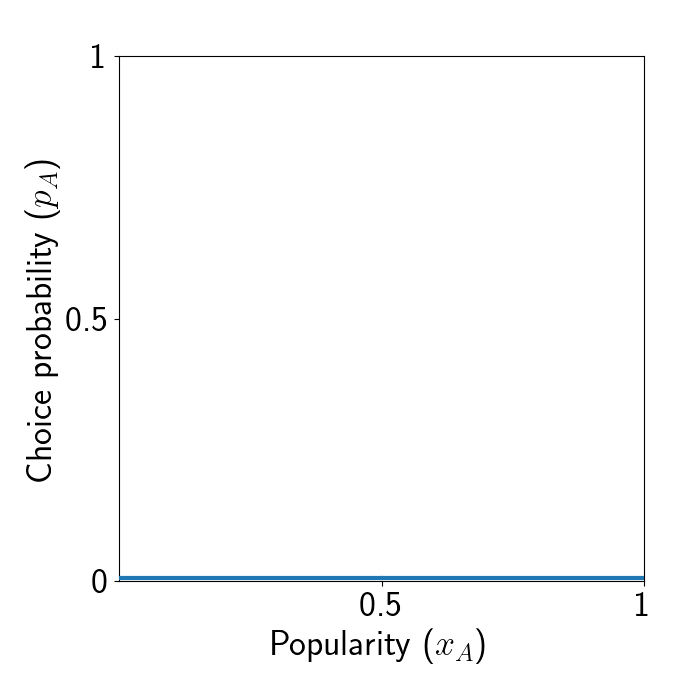}
    \includegraphics[height=0.47\columnwidth, trim = 45 0 0 0,clip]{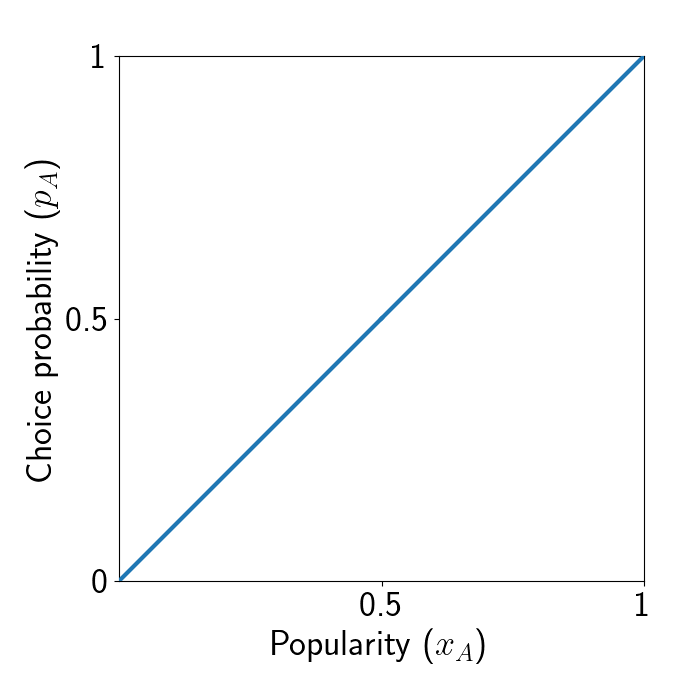}
    \includegraphics[height=0.47\columnwidth, trim = 45 0 0 0,clip]{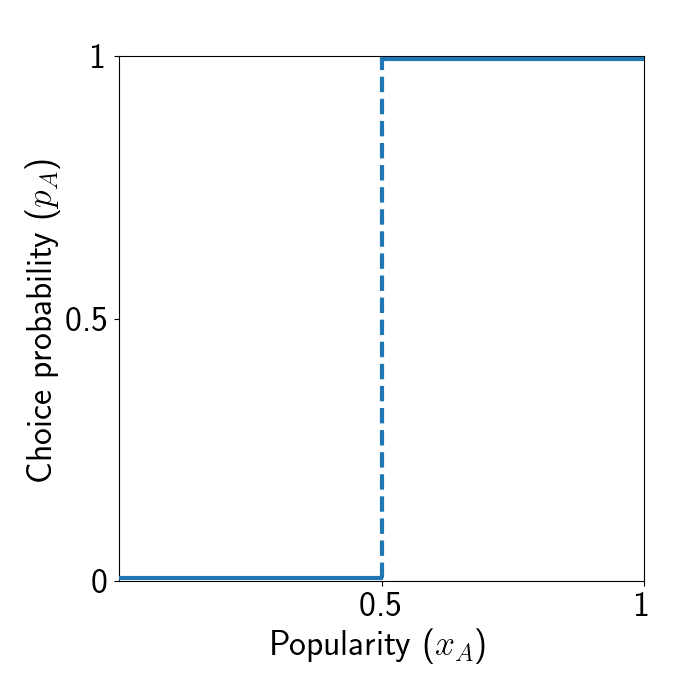}
    \includegraphics[height=0.47\columnwidth]{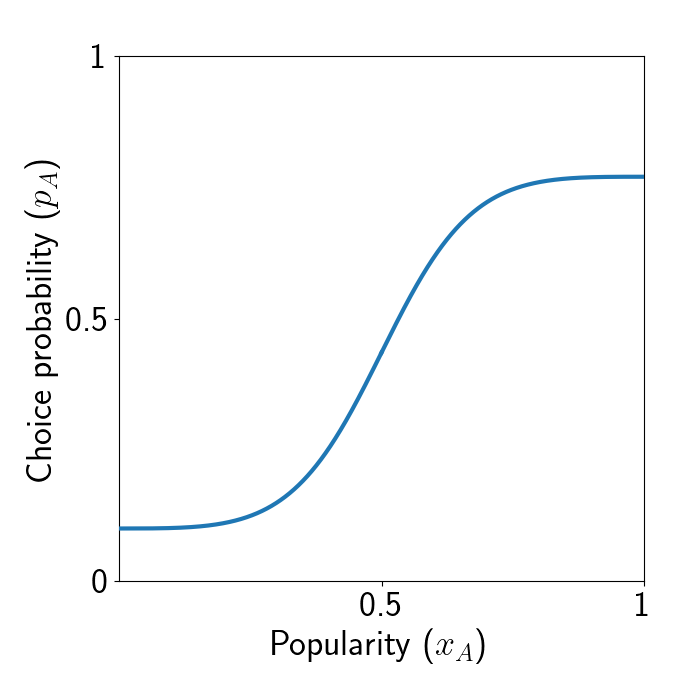}
    \caption{
    Possible individual-level influence curves. \textbf{Top left:} Influence curve describing an individual who never chooses option A, no matter how many others have chosen it. \textbf{Top right} An individual who chooses each option with probability equal to its popularity. \textbf{Bottom left:} An individual following the majority heuristic. \textbf{Bottom right:} An individual whose probability of choosing A is an arbitrary, non-linear function of its popularity.} \label{appendixIndividualInfluenceCurvesFigure}
\end{figure}

Influence curves as used in this paper are an aggregate level theoretical construct, produced by averaging behavioral tendencies and data across individuals. Influence curves, however, can also be conceived at the individual level, expressing the extent to which different individuals are influenced by others in specific contexts and the behavioral strategies they might follow (see \cref{appendixIndividualInfluenceCurvesFigure}). For example, in the quiz experiments reported in FV2021, some people might be certain that they know the answer, which means that their influence curves will be  unresponsive to social influence (\cref{appendixIndividualInfluenceCurvesFigure}, top left). Others may be completely ignorant of the answer and use a proportional response or a follow-the-majority strategy (\cref{appendixIndividualInfluenceCurvesFigure} top right and bottom left, respectively). Yet other individuals might have some background knowledge without being certain about the answer, and thus might be influenced in an arbitrary non-linear way by what other people have responded (\cref{appendixIndividualInfluenceCurvesFigure}, bottom right).

If the influence curves of individuals are known, they can be aggregated in order to derive the population-level influence curve $f(x)$ which can be used to predict the possibility of lock-in, under the assumption of random sampling (see Background and definitions section). For example, Granovetter's famous threshold model \citeyearpar{granovetter1978threshold}, assumes that people's individual level curves are step functions, jumping discontinously from probability 0 to 1 (similar to \cref{appendixIndividualInfluenceCurvesFigure}, bottom right), but that the thresholds in the population vary. That is, some individuals could be triggered to act even if a small proportion of others did, whereas others may require a larger established base of other people acting before they do so themselves. When aggregated, these discontinuous step functions will often produce a smooth, continuously increasing function (e.g., if the thresholds are normally distributed). The aggregate curve will have discontinuities if there is a positive fraction of the population with the same threshold, which may occur, for example, if a subset of the population adopts the follow-the-majority heuristic.

In practice, empirical data often allows to derive directly the aggregate influence curve, but not the individual-level ones. Deriving individual level curves would require empirical observations at different levels of social influence for a single individual in the same exact social influence setting, which in most cases is hard to obtain.

\newpage

\bibliographystyle{apalike}
\bibliography{main}

\renewcommand{\theequation}{S\arabic{equation}}
\renewcommand{\thetable}{S\arabic{table}}
\renewcommand{\thefigure}{S\arabic{figure}}
\setcounter{equation}{0}
\setcounter{table}{0}
\setcounter{figure}{0}
\def\thesection{\arabic{section}}

\title{Supplementary Information for: The marginal majority effect: when social influence produces lock-in}

\maketitle

\section{Figures including data at $x=0.5$}

\begin{figure}
    \centering
    \textbf{\hspace{0.7cm} V2019 - All data}\par\medskip
    \includegraphics[width=0.5\columnwidth, trim =0 0 0 50, clip]{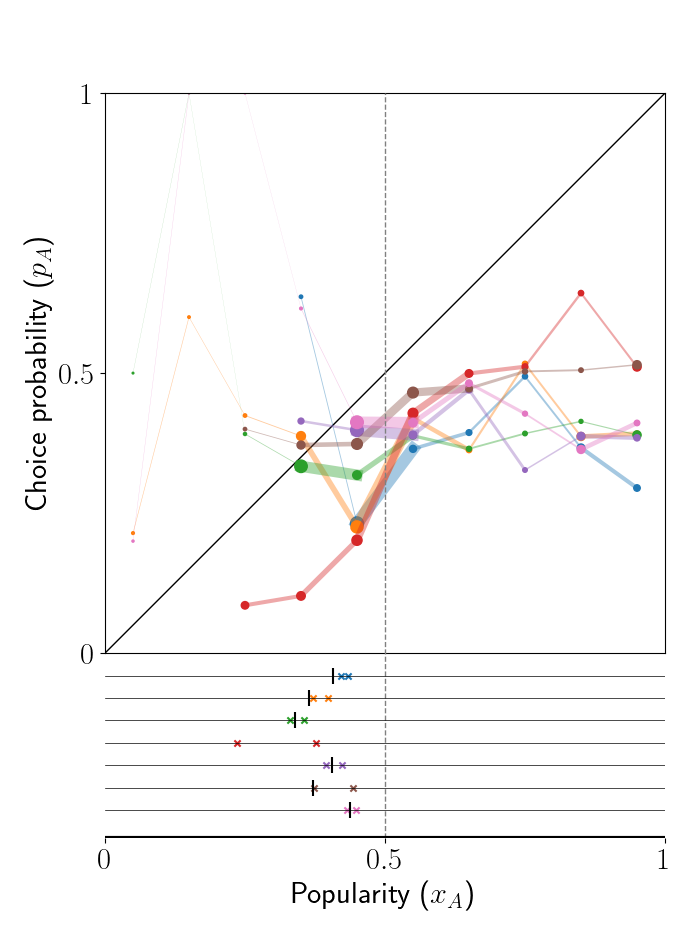}
    \caption{Replication of Fig. 5 of the main text, without removing data with $x=0.5$.}
    \label{fig_vdRijt_with_x=0.5}
\end{figure}

\begin{figure}
    \centering
    \includegraphics[width=\columnwidth]{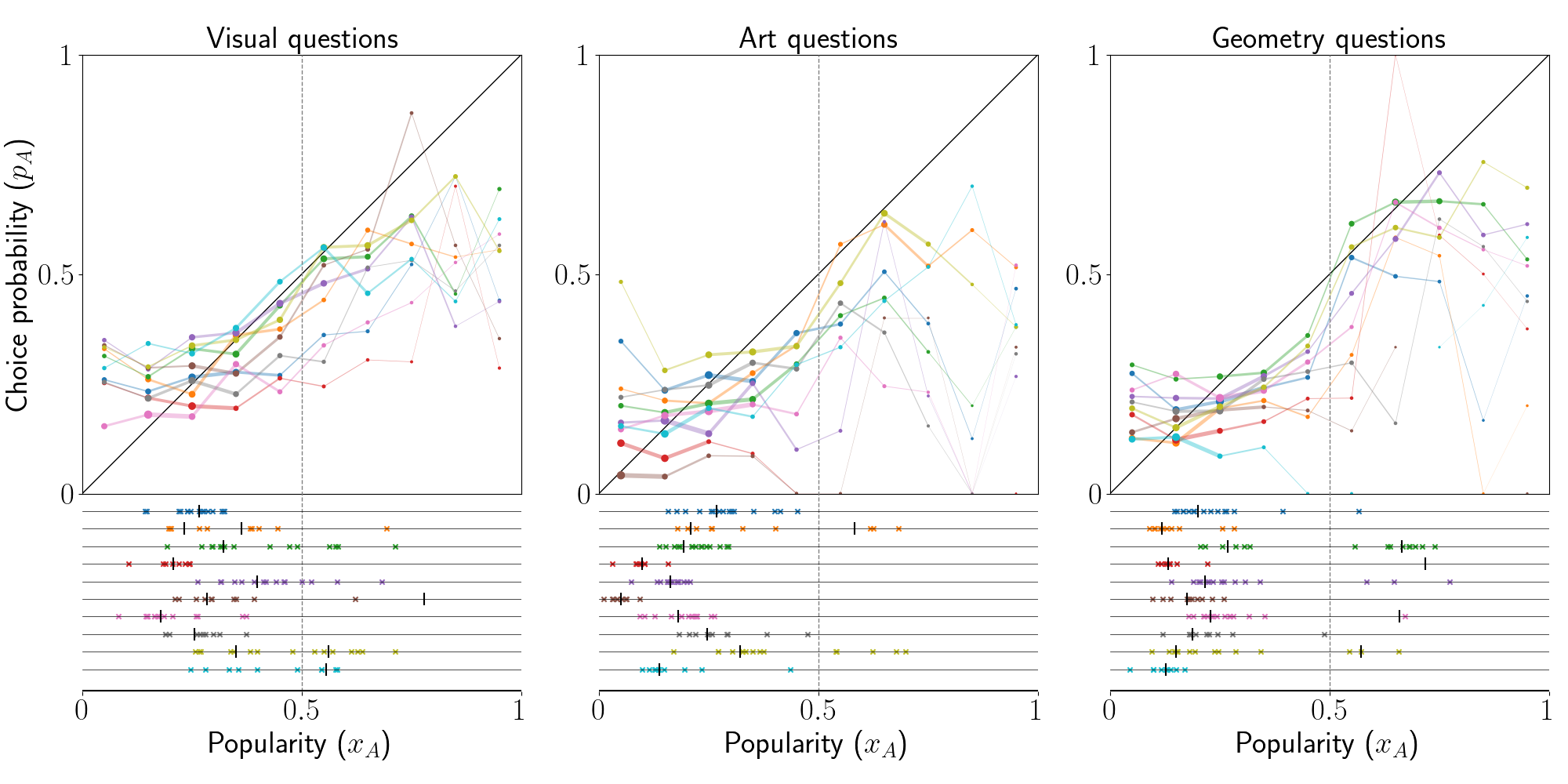}
    \textbf{\hspace{0.7cm} FV2021 - All data}\par\medskip
    \caption{Replication of Fig. 7 of the main text, without removing data with $x=0.5$.}
    \label{fig:Frey-vdRijtSI_with_x=0.5}
\end{figure}

In Figs. 5 and 7 of the main text, points that fell exactly on $x=0.5$ were removed from the graph, in order for the marginal majority effect to be visible (see main text Methods). \Cref{fig_vdRijt_with_x=0.5,fig:Frey-vdRijtSI_with_x=0.5} are replications of those figures without removing those points, but instead treating them the same way we treat other points that fall on bin endpoints, i.e., by placing them with weight 0.5 on each of the adjacent bins. Recall that we refer to the datasets analyzed with the acronyms V2019 \citep{van2019self}, MDRT2019 \citep{macy2019opinion}, and FV2021 \citep{frey2021social}.

\newpage

\section{Out-of-sample prediction of end-of-trial proportions and lock-in}

In Figs. 5-8 of the main text, data from all trials were used both to estimate the influence curves and to check whether lock-in occurred.\footnote{In V2019 a third trial contained external manipulations of the popularity and was used only for influence curve estimation. Also, in FV2021, trials of only 15 participants were used only for influence curve estimation.} To see why this can be problematic, note that in principle we don't know whether the influence curve model is correct. It is conceivable that people do not pay attention to the proportions but instead to some other function of the counts, in a way that the process cannot be described via an influence curve.\footnote{This is not relevant for MDRT2019, because users were not shown the counts of previous answers.} For example, the proportions of people who choose each item need not converge, or they might not always converge to one of a small set of points (e.g., the stable equilibria of $f$). In that case, the proportion estimates for the equilibria would be merely properties of the available data, rather than estimates of an underlying ground truth. In particular, out-of-sample predictions would turn out to be inaccurate.

Here, we provide evidence that the model makes accurate out-of-sample predictions of the end-of-trial proportions and lock-in, which gives further support for the suitability of the model. Specifically, we repeat the analysis reported in the main text, but splitting the trials for each experimental item (question) into two sets, the first of which we use for estimating the influence curve and marginal majority effect (training set) and the second for assessing whether lock-in occurs (test set). Because V2019 included only two trials, we exclude it from the analysis.

\subsection{Splitting the data in training and test sets}
\label{sectionSplit}

For MDRT2019 we use half of the trials (4/8) of each question for training and the rest for testing. For FV2021 the training set consists of half of the trials with 100 participants (5/10 or 8/15 trials, depending on the question) and all of the trials with 15 participants (20 or 30 trials, depending on the question), because the latter are too short to reveal the occurrence of lock-in. The rest of the trials with 100 participants form the test set.

In order to reduce any bias in the choice of how we split the trials in the figures that we include here, we have ordered the trials by their index in the original datasets that are available online and assigned the first half of the eligible trials to the training set. We also provide summary statistics from replications of the analyses with random splits.

\subsection{Results: influence curves and lock-in}

\begin{figure}
    \centering
    \textbf{\hspace{0.7cm} MDRT2019 - Out-of-sample}\par\medskip
    \includegraphics[width=0.5\columnwidth]{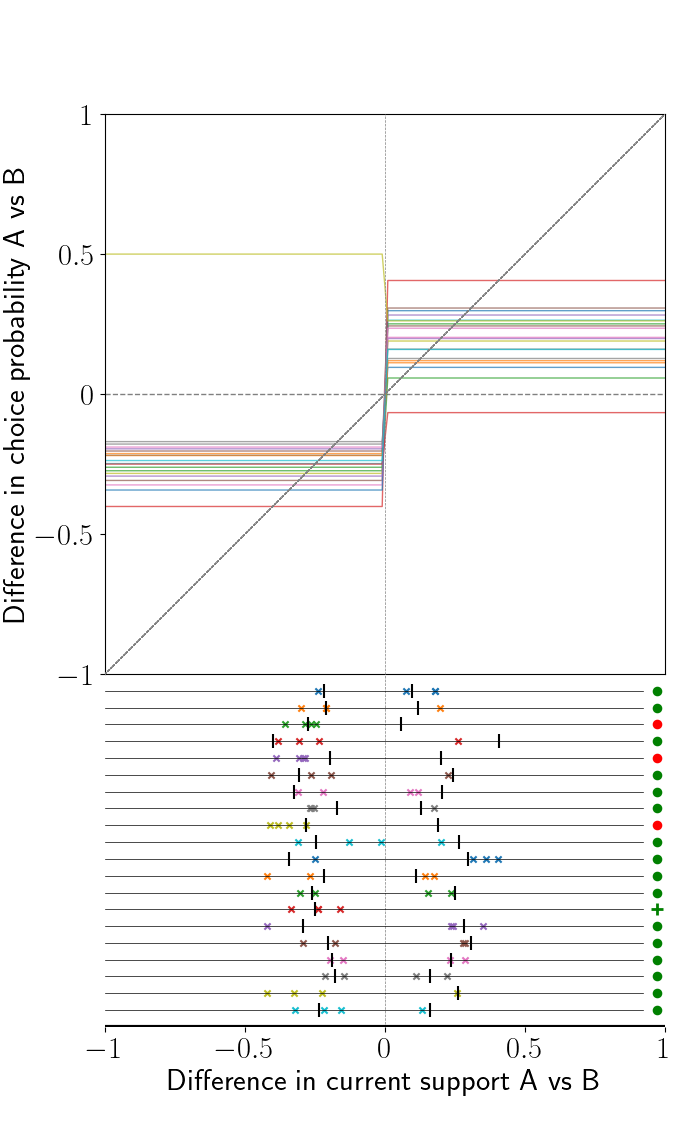}
    \caption{Replication of Fig. 6 of the main text, using a subset of the trials for estimating the influence curves and the rest of the trials for depicting end-of-trial proportions. The marks at the bottom right of the graph indicate for which of the questions the predictions are accurate. \textbf{Circles/crosses:} Lock-in predicted/not predicted based on influence curve entering or not the lock-in region. \textbf{Green/red color:} Prediction accurate/not accurate.}
    \label{figInfluenceMacySI}
\end{figure}

\begin{figure}
    \centering
    \textbf{\hspace{0.5cm} FV2021 - Out-of-sample}\par\medskip
    \includegraphics[width=\columnwidth]{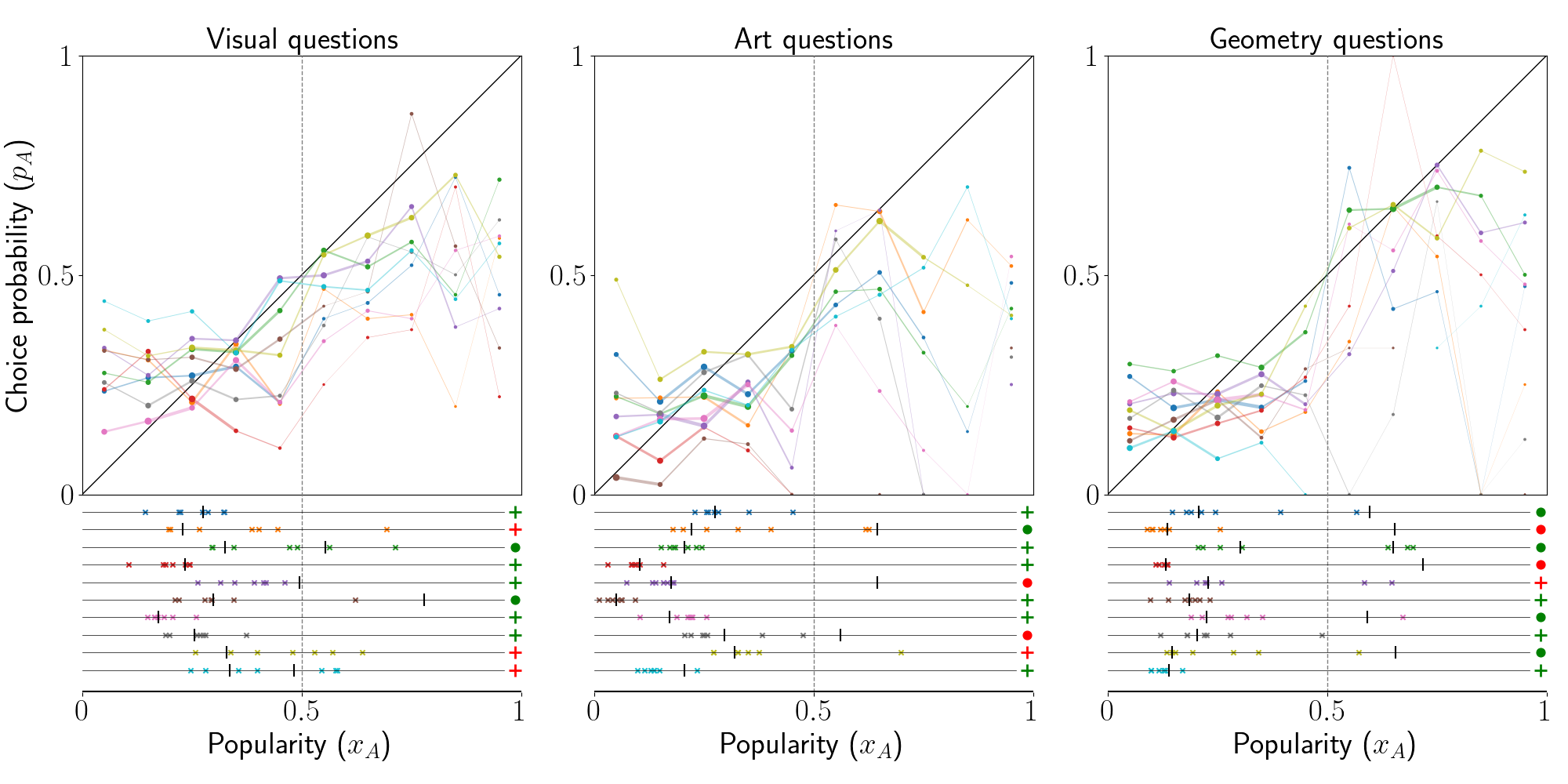}
    \caption{Replication of Fig. 7 of the main text, using a subset of the trials for estimating the influence curves and the rest of the trials for depicting end-of-trial proportions. The marks at the bottom right of each panel are as in \cref{figInfluenceMacySI}.}
    \label{fig:Frey-vdRijtSI}
\end{figure}

\Cref{figInfluenceMacySI,fig:Frey-vdRijtSI} show the influence curves and the predicted equilibria for MDRT2019 and FV2021, respectively, generated from the training set data, and the end-of-trial majorities for the test set trials.

In \cref{figInfluenceMacySI}, there is good agreement between the predicted equilibria and observed end-of-trials proportions. Additionally, like in Fig. 6 of the main text, in all but one cases the possibility of lock-in is predicted. Although only four trials for each question were used to assess the occurrence of lock-in, lock-in was observed in 16/19 cases where it was predicted. The one case where the influence curve is above the $y=0$ line \textit{on the left half of the graph} is due to very small sample size, only 7 answers from Democrats and Republicans together. For all other estimates there were at least 170 answers, and at least 70 from each group (Democrats and Republicans).

In \cref{fig:Frey-vdRijtSI}, the right half of the graph is substantially affected by noise, due to the limited data available. Still, lock-in is observed (end-of-trial majority of option A) in 8 out of 13 cases (62\%) in which it is predicted to be possible, while there are 4 out of 17 cases (24\%) in which lock-in is observed without being predicted. With regards to the location of the end-of-trial proportion, the predictions (downcrossings) are reasonably accurate when option B has an end-of-trial majority (left half of the graph), but less so in the case of lock-in (right half of the graph), again presumably due to the limited amount of data.

\subsubsection*{Robustness check}

The results are similar with different splits of the data into training and test sets. In 100 runs with random split,\footnote{Sizes of test and training sets are as before. In FV2021 all trials of 15 participants are part of the training set.} we find that the influence curves of 28.7 out of 50 items on average entered the lock-in region (aggregate data for MDRT2019 and FV2021), and in 21.7 of those ($75.6\%$) lock-in was observed. In contrast, in only 6.5/21.3 items on average ($30.5\%$) lock-in was observed when the influence curves did not enter the lock-in region. (See main text for a discussion of the possible sources of error.) The confusion matrix is reported in \Cref{table:confusionMatrix}. These results provide support to the suitability of the theoretical model for describing the dynamics of popularity in these studies.

\begin{table}
    \centering
    \caption{Confusion matrix for the lock-in prediction based on whether the influence curve enters the lock-in region. The numbers are averages over 100 runs with random splits of the trials into training and test sets, for MDRT2019 and FV2021 together.\\ }
    \label{table:confusionMatrix}
    \begin{tabular}{c|c|c|c}
         Lock-in & Predicted & Not predicted & \textbf{Total}\\ \hline
         Observed & 21.7 & 6.5 & \textbf{28.2}\\ \hline
         Not Observed & 7 & 14.8 & \textbf{21.8}\\ \hline
         \textbf{Total} & \textbf{28.7} & \textbf{21.3} & \textbf{50}
    \end{tabular}
\end{table}

\subsection{Predicting lock-in from marginal majority effect}

\begin{figure}[hbt]
    \centering
    \includegraphics[width=0.8\columnwidth, trim =0 0 0 0, clip]{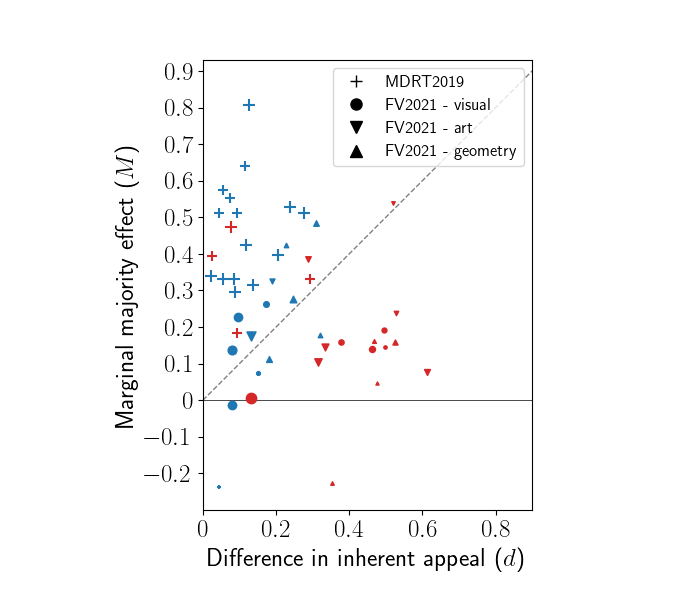}
    \caption{Replication of Fig. 8 of the main text, using a subset of the trials for estimating the marginal majority effect and the rest of the trials for assessing the occurrence of lock-in.}
    \label{fig2}
\end{figure}

\Cref{fig2} is a replication of Fig. 8 of the main text, with out-of-sample prediction, using the training data to estimate the marginal majority effect and the test data to decide whether lock-in is observed.\footnote{The difference in inherent appeal is estimated from the data of the control experiment, similarly to Fig. 8 of the main text.} The split into training and test sets is described in \cref{sectionSplit}. We see that the relation between marginal majority effect and quality difference continues to be a good predictor of the possibility of lock-in. Specifically, lock-in is observed in 23/29 cases (82\%) where it is predicted to be possible ($M>d$), and in 5/18 (28\%) cases when $M<d$ (in which case the theory is inconclusive about its possibility).

\subsubsection*{Robustness check}
The results are similar when the split between training and test sets is random. Specifically, for the two experiments together, in 100 runs with random split, the condition $M>d$ was satisfied for 29.5 items on average (lock-in predicted), and for 24 of those on average lock-in was observed ($81\%$). In contrast, lock-in was observed on average for only 4.1/20.5 ($20\%$) of the items for which $M<d$ (and for which there is no theoretical prediction). The confusion matrix is reported in \Cref{table:confusionMatrix_majority}.

These predictions are more accurate than those based on the influence curve entering the lock-in region, despite using less data.

\begin{table}
    \centering
    \caption{Confusion matrices for the lock-in prediction based on the condition $M>d$. The numbers are averages over 100 runs with random splits of the trials into training and test sets, for MDRT2019 and FV2021 together. Note that when $M\leq d$, there is no theoretical prediction, either for presence or absence of lock-in.\\ }
    \label{table:confusionMatrix_majority}
    \begin{tabular}{c|c|c|c}
         Lock-in & Predicted & No prediction & \textbf{Total}\\ \hline
         Observed & 24 & 4.1 & \textbf{28.1}\\ \hline
         Not Observed & 5.5 & 16.4 & \textbf{21.9}\\ \hline
         \textbf{Total} & \textbf{29.5} & \textbf{20.5} & \textbf{50}
    \end{tabular}
\end{table}

\end{document}